\documentclass{emulateapj}
\slugcomment{{\sc Accepted to ApJ:} August 30, 2011} 
\shorttitle{CO as a Probe of Inner Disk Structure}

\shortauthors{Salyk et al.}
\newcounter{minirefcount}
\newcounter{refcount}
\usepackage{amssymb}
\usepackage{amsmath}
\usepackage{wasysym}

\newcommand{\miniref}[2]{\refstepcounter{minirefcount}\label{#2}(\arabic{minirefcount}) #1}

\begin{document}

\title{CO rovibrational emission as a probe of inner disk structure}
\author{C. Salyk}
\affil{McDonald Observatory, The University of Texas at Austin, 1 University Station, C1402, Austin, TX 78712, USA}
\author{G.~A. Blake}
\affil{Division of Geological \& Planetary Sciences, Mail Code 150-21, California Institute of Technology, Pasadena, CA 91125, USA}
\author{A.~C.~A. Boogert}
\affil{California Institute of Technology, Infrared Processing and Analysis Center, Mail Code 100-22, Pasadena, CA 91125, USA}
\author{J.~M. Brown}
\affil{Harvard-Smithsonian Center for Astrophysics, 60 Garden Street, Cambridge, MA 02138, USA}

\begin{abstract}
We present an analysis of CO emission lines from a sample of T Tauri, Herbig Ae/Be, and transitional disks with known inclinations, in order
to study the structure of inner disk molecular gas.  We calculate CO inner radii by fitting line profiles with a simple parameterized model.  
We find that, for optically thick disks, CO 
inner radii are strongly correlated with the total system luminosity (stellar plus accretion), and consistent with the dust sublimation radius.  
Transitional disk inner radii show the same trend with luminosity, but are systematically larger.  Using rotation diagram fits, we
derive, for classical T Tauri disks, emitting areas consistent with a ring of width $\sim$0.15 AU located at the CO inner radius; emitting areas
for transitional disks are systematically smaller.  We also measure lower rotational temperatures for transitional disks, and disks around Herbig Ae/Be stars, 
than for those around T Tauri stars.  Finally, we find that rotational temperatures are similar to, or slightly lower than, the expected temperature
of blackbody grains located at the CO inner radius, in contrast to expectations of thermal decoupling between gas and dust.
\end{abstract}

\keywords{protoplanetary disks; stars: pre-main-sequence}
\section{Introduction}
Observations of the inner regions ($\lesssim$ 5 AU) of circumstellar disks are essential for obtaining a complete understanding of star and planet formation.  The inner disk interacts with the central star, thereby controlling the accretion and ejection of material and setting the timescale for star formation and disk evolution.  In addition, inner disks are the likely birthplace of terrestrial planets. However, these regions are difficult to study, because of their small angular size at the distance of nearby star-forming regions and the proximity to their parent stars.

Recently, significant advancements in our understanding of inner disks have been made with two complementary techniques: IR interferometry and high-resolution spectroscopy.  IR interferometry is capable of observing thermal emission from the dusty component of disks with resolutions of a few mas (probing size scales down to a few hundredths of an AU; Millan-Gabet et al.\ 1999; Eisner et al.\ 2003; Akeson et al.\ 2005b, Monnier et al.\ 2005).   These observations have shown that standard, optically thick accretion disk models (such as outlined by \citealt{Hillenbrand92}) fail to simultaneously fit  spectral energy distributions (SEDs) and visibilities \citep{Millan-Gabet01, Akeson02}.  Better fits to visibilities were obtained with an optically thin inner gas disk interior to a hot, puffed-up wall of dust \citep{Dullemond01, Natta01}, with the dust wall located where temperatures are high enough to sublimate silicate grains \citep{Monnier05}.  Recent observations and modeling efforts have introduced complications to this simple picture, including a significant source of emission of unknown origin {\it inside} the dust sublimation radius \citep[for an extensive review, see][]{Dullemond10}.  However, the major conclusion --- that there is a defined inner radius consistent with dust sublimation --- still holds true.

The complementary technique of high-resolution spectroscopy has been used to study rovibrational emission from hot gaseous molecules, specifically H$_2$O and CO \citep[e.g.,][]{Brittain03,Najita03, Blake04, Carr04, Rettig04, Salyk08}, that originate in the inner disk atmosphere. Because of the high temperatures ($\gtrsim 1000$ K) required to populate vibrationally excited states,  CO vibrational emission originates at disk radii similar to those probed by IR interferometry ($\lesssim$ a few AU). When the spectral resolution is high enough that the emission lines are spectrally resolved, the line profile acts as a proxy for the spatial location of the gas, under the assumption that the emission originates in a rotating Keplerian disk.  Since the emission line wings represent the highest observed velocities, the flux in the line wings originates at the inner edge of the molecular disk.  Thus, emission lines (after a correction for disk inclination) can be used to measure molecular gas inner radii.

In this work, we present observations of CO $v$=1$\rightarrow$0 emission from a large sample of protoplanetary disks, including ``classical'' optically thick disks around T Tauri stars (cTTs's) and Herbig Ae/Be (HAeBe) stars, as well as transitional disks --- disks whose inner regions are depleted of small dust grains \citep[e.g.,][]{Koerner93, Calvet02}.  The disks in the chosen sample all have known inclinations, allowing us to
use line profiles to derive CO inner radii, which we then compare to the interferometric dust inner radii.  We also fit the emission line fluxes with a local thermodynamic equilibrum (LTE) slab emission model, and relate the bulk properties of the gas to the measured CO inner radii.

\section{Observations and Reduction}
Most of our spectra were obtained with NIRSPEC \citep{McLean98}, a high-resolution ($R\sim$25,000, FWHM$\sim$12.5 km s$^{-1}$) spectrometer on the Keck II telescope.  These observations are derived from several observing runs spanning the years 2001--2010, and are part of a large NIRSPEC survey of protoplanetary disks, portions of which have been previously presented \citep{Blake04, Salyk07, Salyk09, Salyk11}.  The data were observed in the $M$ band in echelle mode with a 0.$''$43$\times$24$''$ slit.  Each source was observed in at least two grating settings, thereby encompassing wavelengths between $\sim4.65$ and $5.15\,\mu$m, with the exception of a hole between orders at $\sim4.8$--$4.95\,\mu$m.  This wavelength range covers the first two R-branch lines and the low/mid P-branch ($J=1-12$ and $J=30-40$) of the $v$=1$\rightarrow$0 CO rovibrational spectrum, as well as the H I Pf$\beta$ and Hu$\varepsilon$ transitions, and in a few cases the Hu$\delta$ transition.  An observation log is provided in Table \ref{tbl:log}.

TW Hya's CO emission lines were not well resolved with NIRSPEC and this source was therefore observed with Phoenix \citep{Hinkle03} on Gemini South, as first reported in \citet{Salyk07}.  It was observed on 2006 April 7 and 8 using the 0.$''$35$\times$14$''$ slit.  With its significantly higher spectral resolution ($R$ $\sim$ 60,000, FWHM $\sim$ 5 km  s$^{-1}$), but smaller spectral coverage, the Phoenix observations resolve
the emission lines from P(6) to P(9).

Objects were observed in nod pairs and then differenced.  Exposure times (integration time multiplied by co-adds) were limited to one minute in length to minimize atmospheric changes between nods.  Wavelengths were calibrated using telluric emission lines.  Nearby A and B standard stars were observed to correct for telluric absorption features.  Any H I lines present in the standard star spectra were fitted by Kurucz models before dividing source spectra by the standard.  Standard stars were also utilized for flux calibration, using $M$-band fluxes estimated from Two Micron All Sky Survey $K$-band photometry \citep{Skrutskie06} and spectral types from the literature.  Portions of the spectra with poor atmospheric transmission (typically $\lesssim 70\%$, but the exact percentage for each source was determined empirically) were removed.  Wavelengths were shifted to correct for a Doppler shift due to Earth's motion, which depends on the time of year.  Many sources were observed at multiple Doppler shifts, thereby ``filling in'' regions of poor atmospheric transmission and creating complete line profiles, even for low-excitation lines.  (High excitation CO lines do not suffer greatly from telluric absorption).  

A more detailed explanation of the data acquisition and reduction used for the NIRSPEC observations in our survey can be found in \citet{Salyk09}.

\begin{deluxetable*}{ll}

\tablecaption{Observation Log \label{tbl:log}}
\tabletypesize{\scriptsize}
\tablehead{Name & Dates}
\startdata
AA Tau & 2003 Nov  3, 2004 Dec 27, 2004 Dec 30\\
AB Aur & 2001 Jan 30, 2001 Aug  7--8,  2002 Jan  3,  2002 Dec 18\\
AS 205 N & 2002 Apr 21, 2002 Jul 22\\
BP Tau & 2009 Oct 11\\
DF Tau & 2005 Dec 21\\
DL Tau & 2003 Nov  3, 2007 Dec 26\\
DoAr 44 & 2002 Jul 22, 2004 Jul 23--24, 2005 Apr 26, 2010 Apr 22\\
DO Tau & 2005 Dec 18\\
DR Tau & 2002 Dec 17--18, 2005 Dec 18\\
GG Tau & 2002 Jan  3, 2002 Dec 17--18, 2003 Nov  2--3\\
GK Tau & 2009 Dec 27--28\\
GM Aur & 2004 Dec 27\\
GSS 39 & 2009 Jul 13, 2010 Apr 22\\
HD 135344 B & 2005 Apr 24, 2006 Jul 6\\
HD 141569 A & 2002 Apr 21, 2008 Jul 9\\
HD 150193 & 2002 Jul 22, 2007 Mar 6\\
HD 163296 & 2001 Aug  6, 2001 Aug  8, 2002 Apr 21, 2002 Jul 22\\
HD 190073 &2002 Jul 22, 2005 Sep 22\\
HL Tau & 2001 Oct 25, 2002 Dec 17--18\\
LkH$\alpha$ 330 & 2002 Dec 16, 2002 Dec 18, 2003 Nov 2--3, 2004 Dec 27, 2004 Dec 29\\
MWC 480& 2001 Jan 30, 2001 Aug  8, 2002 Jan  3, 2002 Dec 17--18\\
MWC 758 & 2002 Apr 21, 2002 Dec 16, 2002 Dec 18\\
SR 9 & 2004 Jul 23--24\\
SR 21 & 2002 Apr 21, 2006 Jul 6--7, 2007 Mar 06\\
SU Aur & 2002 Dec 16, 2002 Dec 18, 2003 Nov 2--3\\
T Tau & 2002 Jan 3, 2002 Dec 17, 2004 Dec 27, 2008 Dec 10\\
TW Hya & 2002 Dec 18-20, 2004 Dec 27, 2005 Dec, 2005 Apr 24, 2006 Apr 7--8 \tablenotemark{a}\\
UX Tau A & 2006 Dec 28--29, 2007 Oct 29--30\\
VV Ser & 2001 Aug  6,  2003 Jul  9--11, 2004 Jul 23--24\\
V1121 Oph & 2002 Jul 21, 2008 Apr 18\\
Wa Oph 6 & 2005 Apr 26\\
\enddata
\tablenotetext{a}{2006 April observations obtained with Phoenix on Gemini South.}
\end{deluxetable*}

\section{Sources and Spectra}
The sample analyzed in this work consists of 32 young stars with circumstellar disks, including cTTs's, HAeBe stars, and stars with transitional disks (31 from our own survey, and one obtained from Najita et al.\ 2003).  A description of the sample can be found in Table \ref{tbl:params}.  This particular subset of our large survey was chosen for analysis based on the following characteristics --- CO in emission, with no large line-shape asymmetries, and known disk inclination --- with the goal of measuring CO inner radii from the line profiles.  
(Note that only two stars with disks --- SR 24 and VSSG 1 --- were eliminated due to line asymmetries, so such asymmetries are not characteristic of typical disks).  Our sample spans a large range of stellar masses and more than two orders of magnitude in luminosity, making this the first study capable of exploring and comparing inner gas radii for a large range of luminosities and stellar types.  In addition, our sample includes nine transitional disks (DoAr 44, GM Aur, HD 135344 B, HD 141569 A, LkCa 15, LkH$\alpha$ 330, SR 21, TW Hya and UX Tau A), making it possible to systematically compare their inner radii with classical disks.

\begin{figure*}
\includegraphics[angle=0, scale=0.9]{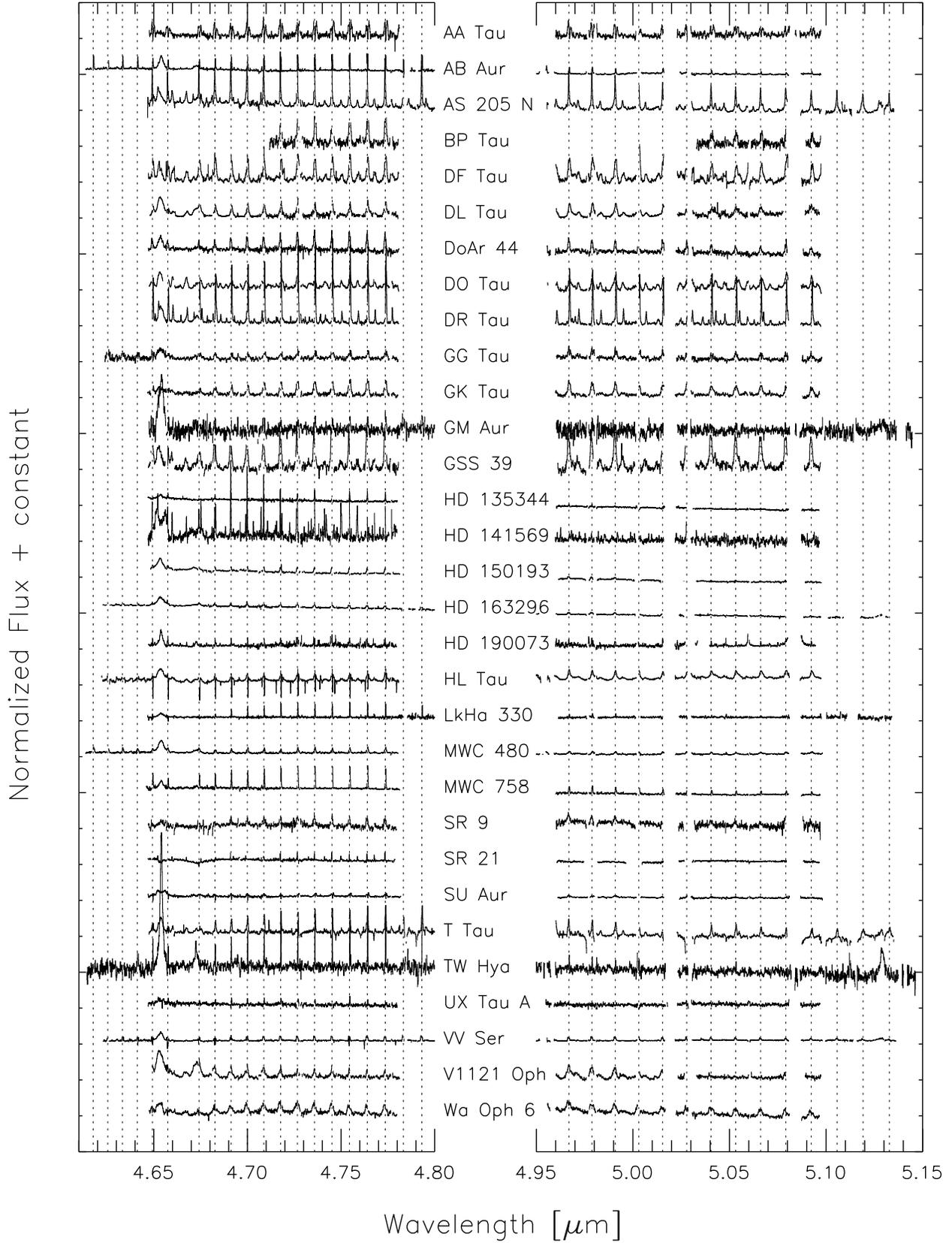}
\caption{Complete set of NIRSPEC spectra.  Dashed vertical lines mark CO P- and R-branch lines. 4.654, 4.673, and 5.129 $\mu$m emission lines are HI Pf$\beta$, Hu$\varepsilon$, and Hu$\delta$, respectively. Additional features present in some spectra include $^{13}$CO 1$\rightarrow$0 and $^{12}$CO 2$\rightarrow$1 lines.
 \label{fig:spec}}
 \end{figure*}

The complete set of spectra is shown in Figure \ref{fig:spec}, with CO $v=1\rightarrow0$ emission lines marked with dotted lines.  Note that every source shows CO P- and R-branch emission lines (and most also show between one and three H I emission lines), but the spectra display a variety of line/continuum ratios and excitation temperatures (as reflected in the ratio of low- to high-excitation line strengths).

\begin{deluxetable*}{lccccclcccc}
\tablecaption{Stellar Parameters \label{tbl:params}}
\tablewidth{0pt}
\tabletypesize{\scriptsize}
\tablehead{\colhead{Star}  &  \colhead{$M_\star$}  &  \colhead{$L_\star$} & \colhead{$d$} & \colhead{$i$} & \colhead{$r_\mathrm{dust}$} & \colhead{log($\dot{M}$)} & 
\colhead{$v\,\mathrm{sin}(i)$}&Type\tablenotemark{a}&Refs\\
&\colhead{($M_\odot$)}&\colhead{($L_\odot$)}&\colhead{(pc)}&\colhead{($^\circ$)}&\colhead{(AU)}&\colhead{($M_\odot$ yr$^{-1}$)}&\colhead{(km s$^{-1}$)}\\ }  
\startdata
    AA Tau &   0.67 &   0.98  &   140  &         75$\pm$ 10  &                \nodata  &   $-8.2$  &   11.4  & TT &               \ref{Bec90},\ref{Cla00},\ref{Joh02},\ref{Kit02},\ref{Wic98} \\
    AB Aur &   2.4 &  47  &   144  &         21$\pm$ 0  &          0.30$\pm$0.01  &   $-5.8$  &   80  & H& \ref{Cor05},\ref{Eis07a},\ref{Hil92},\ref{Pog05},\ref{Tan08} \\
  AS 205 A &   1.0 &   4.0  &   125  &         25$\pm$10  &          0.18$\pm$0.01  &   $-7.1$  &   14.9  &TT &     \ref{And09},\ref{Eis05},\ref{Eis09} \\
    BP Tau &   0.77 &   0.83  &   140  &  30$_{- 2}^{+ 4}$  &          0.08$\pm$0.05  &   $-7.9$  &    7.8  &TT &\ref{Ake05b},\ref{Cla00},\ref{Kit02} \\
    DF Tau &   0.53 &   2.97  &   140  &  78$_{-35}^{+12}$  &                \nodata  &   $-6.7$  &   16.1  & TT &               \ref{Bec90},\ref{Bou90},\ref{Cla00},\ref{Joh02},\ref{Wic98}\\
    DL Tau &   0.56 &   0.77  &   140  &         25$\pm$ 5  &                \nodata  &   $-7.6$  &   16  &  TT &              \ref{Bec90},\ref{Cla00},\ref{Joh02},\ref{Kit02},\ref{Wic98}\\
   DoAr 44 &   1.4 &   1.3  &   125  &         45$\pm$10  &                \nodata  &\nodata  &\nodata  &TT, Tr&                                    \ref{And09},\ref{Lah07}\\
    DO Tau &   0.72 &   1.38  &   140  &         42$\pm$ 4  &                \nodata  &   $-7.5$  &   11.1  & TT&              \ref{Bec90},\ref{Cla00},\ref{Kit02},\ref{Whi01},\ref{Wic98} \\
    DR Tau &   0.4 &   3.00  &   140  &         37$\pm$ 3  &          0.07$\pm$0.03  &   $-5.1$  &   $\leq10$  &TT& \ref{Ake05a},\ref{Ake05b},\ref{Cla00},\ref{Ise09},\ref{Joh02}\\
    GG Tau A &   0.73 &   1.50  &   140  &         37$\pm$ 1  &                \nodata  &   $-7.5$  &   10.2  & TT&          \ref{Cla00}, \ref{Joh02},\ref{Sim00},\ref{Wic98} \\
    GK Tau &   0.75 &   1.17  &   140  &         52$\pm$10  &                \nodata  &   $-9.3$  &   18.7  &  TT&      \ref{Bou95},\ref{Cla00},\ref{Joh02}, \ref{Wic98} \\
    GM Aur &   0.5 &   0.74  &   140  &         51$\pm$ 2  & 0.22$_{-0.09}^{+0.08}$  &   $-8.2$  &   12.4 &TT, Tr &\ref{Ake05b},\ref{Cla00},\ref{Ise09} \\
    GSS 39 &   0.6 &   1.0  &   125  &         46$\pm$ 7  &                \nodata  &   $-7.2$  &\nodata  &   TT&                  \ref{And09},\ref{Lah07}\\
 HD 135344 B &   1.8 &   6.8  &    84  &         21$\pm$10  &                \nodata  &\nodata  &\nodata  & H, Tr&                                 \ref{Bro07},\ref{Bro09} \\
 HD 141569 A &   2.00 &  25.77  & 108  &         51$\pm$ 3  &                \nodata  &  $-11.0$  &  258  &    H, Tr&                        \ref{Mer04},\ref{Mor01},\ref{Wei99}\\
 HD 150193 &   2.3 &   1.47  &   150  &         38$\pm$ 9  & 0.58$_{-0.09}^{+0.15}$  &   $-6.2$  &  100  &H&   \ref{Fuk03},\ref{Hil92},\ref{Mon05},\ref{van98}\\
 HD 163296 &   2.3 &  36.0  &   122  &         51$\pm$ 2  &          0.28$\pm$0.01  &   $-7.1$  &  120  &  H&   \ref{Eis09},\ref{Ise09},\ref{Pog05} \\
 HD 190073 &   5.05 & 470.8  &   767  &  28$_{- 8}^{+ 7}$  &          0.62$\pm$0.01  &\nodata  &   15  &H&         \ref{Cut90},\ref{Eis09},\ref{Mon09},\ref{Pog05} \\
    HL Tau &   0.55 &   0.9  &   140  &         53$\pm$ 1  &                \nodata  &   $-8.8$  &\nodata  &   TT&                      \ref{Bec90},\ref{Kit02},\ref{Whi01},\ref{Wic98}\\
   LkCa 15 &   0.7 &   0.74  &   140  &         58$\pm$ 4  & 0.10$_{-0.04}^{+0.03}$  &  $-8.8$  &   12.5  &H, Tr&  \ref{Ake05b},\ref{Cla00},\ref{Ise09} \\
  LkHa 330 &   2.5 &  16  &   250  &         42$\pm$10  &                \nodata  &   $-8.8$  &\nodata  &  H, Tr&                     \ref{Bro07},\ref{Bro09},\ref{Sal09} \\
   MWC 480 &   1.65 &  11.5  &   140  &         26$\pm$ 7  &          0.28$\pm$0.01  &\nodata  &   85  &H&                \ref{Cha08},\ref{Eis09},\ref{Pog05},\ref{Sim00} \\
   MWC 758 &   1.80 &  11  &   140  &         16$\pm$ 4  &          0.35$\pm$0.03  &\nodata  &\nodata  &H&                 \ref{Cha08},\ref{Eis09}\\
      SR 9 &   1.2 &   2.7  &   160  &         34$\pm$10  & 0.23$_{-0.10}^{+0.11}$  &   $-7.5$  &   15.2  &TT&      \ref{Bou90},\ref{Eis05}\\
     SR 21 &   2.0 &  11  &   125  &         22$\pm$10  &                \nodata  &\nodata  &\nodata  & TT, Tr&                      \ref{And09},\ref{Lah07}\\
    SU Aur &   1.97 &  10.70  &   140  &         52$\pm$10  &          0.18$\pm$0.03  &   $-8.2$  &   65.0  &TT&  \ref{Ake05a},\ref{Cal04},\ref{Cla00} \\
     T Tau &   2.41 &   8.91  &   140  &  29$_{-15}^{+10}$  &          0.22$\pm$0.05  &   $-7.5$  &   20.1  & TT&    \ref{Ake02},\ref{Cla00},\ref{Joh02}\\
    TW Hya &   0.7 &   0.25  &    51  &          7$\pm$ 1  &          0.06$\pm$0.01  &   $-9.4$  &    4  & TT, Tr&    \ref{del04},\ref{Eis06},\ref{Ise09},\ref{Qi04} \\
    UX Tau A &   1.1 &   1.0  &   160  &         29$\pm$10  &                \nodata  &   $-9.0$  &   25.4  &  TT, Tr&              \ref{Bro07},\ref{Cla00},\ref{Joh98},\ref{Naj07},\ref{Wic98} \\
    VV Ser &   2.6 &  49  &   260  &         70$\pm$ 5  &          0.59$\pm$0.07  &   $-5.2$  &  200  &H& \ref{Eis07a},\ref{Hil92},\ref{Pon07} \\
 V1121 Oph &   0.9 &   1.5  &   125  &         38$\pm$10  &                \nodata  &   $-7.0$  &\nodata  &  TT&                          \ref{And09},\ref{Lah07}\\
  Wa Oph 6 &   0.9 &   2.9  &   125  &         39$\pm$10  &                \nodata  &   $-7.0$  &   22.9  &  TT&             \ref{And09},\ref{Eis05},\ref{Lah07} \\
\enddata

\tablerefs{\usecounter{refcount}
               \miniref{Akeson et al.\ 2002}{Ake02}
           \miniref{Akeson et al.\ 2005b}{Ake05a}
           \miniref{Akeson et al.\ 2005a}{Ake05b}
             \miniref{Andrews et al.\ 2009}{And09}
           \miniref{Beckwith et al.\ 1990}{Bec90}
               \miniref{Bouvier 1990}{Bou90}
           \miniref{Bouvier et al.\ 1995}{Bou95}
               \miniref{Brown et al.\ 2007}{Bro07}
               \miniref{Brown et al.\ 2009}{Bro09}
              \miniref{Calvet et al.\ 2004}{Cal04}
               \miniref{Chapillon et al.\ 2008}{Cha08}
           \miniref{Clarke \& Bouvier 2000}{Cla00}
           \miniref{Corder et al. 2005}{Cor05}
           \miniref{Cuttela \& Ringuelet 1990}{Cut90}
           \miniref{de la Reza \& Pinzon 2004}{del04}
                       \miniref{Eisner et al.\ 2005}{Eis05}
             \miniref{Eisner et al.\ 2006}{Eis06}
              \miniref{Eisner et al.\ 2007}{Eis07a}
              \miniref{Eisner et al.\ 2009}{Eis09}
              \miniref{Fukagawa et al.\ 2003}{Fuk03}
               \miniref{Hillenbrand et al.\ 1992}{Hil92}
               \miniref{Isella et al. 2009}{Ise09}
               \miniref{Johns-Krull et al.\ 1998}{Joh98}
               \miniref{Johns-Krull \& Gafford 2002}{Joh02}
               \miniref{Kitamura et al.\ 2002}{Kit02}
               \miniref{Lahuis et al.\ 2007}{Lah07}
               \miniref{Mer\'{i}n et al.\ 2004}{Mer04}
               \miniref{Monnier et al.\ 2005}{Mon05}
               \miniref{Montesinos et al.\ 2009}{Mon09}
               \miniref{Mora et al.\ 2001}{Mor01}
               \miniref{Najita et al.\ 2007}{Naj07}
                \miniref{Pogodin et al.\ 2005}{Pog05}
                \miniref{Pontoppidan et al.\ 2007}{Pon07}
                \miniref{Qi et al. 2004}{Qi04}
                \miniref{Salyk et al.\ 2009}{Sal09}
                \miniref{Simon et al.\ 2000}{Sim00}
                 \miniref{Tannirkulam et al.\ 2008}{Tan08}
                 \miniref{van den Ancker et al.\ 1998}{van98}
                 \miniref{Weinberger et al.\ 1999}{Wei99}
                 \miniref{White \& Ghez 2001}{Whi01}  
                \miniref{Wichmann et al.\ 1998}{Wic98}       }
\tablenotetext{a}{Classification used in this work.  H:Herbig Ae/Be, TT:T Tauri and Tr:Transitional.}
\end{deluxetable*}

\section{Analysis of line profiles \label{sec:lines}}
 
\subsection{Constructing Line Composites \label{construction}}
To increase signal-to-noise ratio (S/N) and obtain good velocity coverage for line profiles, we constructed and analyzed line-profile composites.  Because some of the lower-excitation CO lines can be contaminated by foreground absorption (e.g., see HL Tau in Figure \ref{fig:spec}), and because lower-excitation lines are more affected by telluric absorption, we created the line composites from higher-excitation ($J_\mathrm{up}>$20) lines only, whenever possible.  The use of high-excitation lines also has the advantage that, since they originate from only the hottest gas, the inner edge of the disk makes a 
relatively large contribution to the total line flux.  However, when high-excitation lines were not available, we used others.  There is sometimes, but not always, an increase in line width with excitation, which could potentially introduce biases into the analysis; we will discuss this in detail in Section \ref{sec:errors}.  The set of lines used to construct each composite is listed in Table \ref{tbl:widths}, along with the measured line width.

The procedure for creating line composites was as follows.  Observed lines were screened for contamination (from $^{13}$CO and $v=2\rightarrow0$ emission), and contaminated lines were eliminated.  Remaining lines were centered at the theoretical line center \citep{Rothman92}, interpolated on a 5 (3 for TW Hya) km s$^{-1}$ grid, and then averaged.  Line composites are plotted in Figure \ref{fig:fluxfit_plot1}. 

About half of the line composites are single-peaked, while half have evidence for somewhat double-peaked profiles.  Two source composites, AA Tau and SU Aur, have strong central depressions that are likely inconsistent with simple disk emission from a Keplerian disk, with the latter, unfortunately, having no observable lines with $J_\mathrm{up}>$20.  Our sample also includes a few sources known to be strongly centrally peaked (AS 205 N and DR Tau; Bast et al.\ 2011), with the overall line shape consistent with the sum of a Keplerian profile and a slow disk wind \citep{Pontoppidan11}.  We do not make an effort to model or understand these exact structures, and instead focus here on a derivation of inner radii from the line wings.  For more in-depth studies of overall line shapes, utilizing higher-resolution data from
VLT-CRIRES, we direct the reader to \citet{Pontoppidan11},  \citet{Bast11}, and J.M.\ Brown et al.\ (2011, in preparation).

\begin{deluxetable*}{lllll}
\tablecaption{Line Profile Parameters \label{tbl:widths}}
\tabletypesize{\scriptsize}
\tablehead{Star & Lines in Composite & FWHM (km s$^{-1}$) & $R_\mathrm{gauss}$ (AU)\tablenotemark{a} \\}
\startdata
                                                 AA Tau&P(30,31,32,37,38,39,40)& 92& 0.09\\
                                  AB Aur&P(26,27,28,29,30,31,32,34,36,37,38,39)& 28& 0.59\\
                       AS 205 N&P(22,30,31,32,33,34,36,37,38,39,40,41,42,43,45)& 43& 0.12\\
                                                             BP Tau&P(36,37,38)& 87& 0.03\\
                                                 DF Tau&P(30,31,32,36,37,38,40)& 79& 0.10\\
                                                             DL Tau&P(30,31,32)& 87& 0.02\\
                                          DoAr 44&P(30,31,32,34,35,36,37,38,40)& 61& 0.24\\
                                              DO Tau&P(30,31,32,33,36,38,39,40)& 87& 0.05\\
                                        DR Tau&P(30,31,32,33,34,36,37,38,39,40)& 29& 0.27\\
                                        GG Tau A&P(30,31,32,33,34,36,37,38,39,40)& 68& 0.07\\
                                              GK Tau&P(30,31,32,33,36,37,38,40)& 97& 0.06\\
                                                       GM Aur&P(9,10,11,12,14)& 47& 0.19\\
                                                    GSS 39&P(30,31,32,37,38,40)& 80& 0.06\\
                                  HD 135344 B&P(1,2,3,5,6,7,8,9,10,11,12)& 19& 1.49\\
                                  HD 141569 A&P(1,2,3,4,5,6,7,8),R(1,0)& 24& 3.68\\
                                                       HD 150193&P(30,32,36,37)& 53& 0.40\\
                                                 HD 163296&P(27,30,31,32,37,38)& 83& 0.26\\
                                              HD 190073&P(30,31,32,34,36,37,38)& 24& 3.38\\
                                     HL Tau&P(27,30,31,32,33,34,36,37,38,39,40)& 96& 0.05\\
                                                        LkHa 330&P(30,31,32,33)& 28& 2.10\\
                                    MWC 480&P(26,27,28,29,30,31,32,36,37,38,40)& 72& 0.08\\
                                                MWC 758&P(30,31,32,33,36,37,38)& 32& 0.19\\
                                                            SR 9&P(30,31,32,37)& 77& 0.08\\
                                               SR 21&P(5,6,7,8,9,10,11,12)& 18& 1.98\\
                                                             SU Aur&P(30,31,32)&121& 0.10\\
                                            T Tau&P(30,31,32,33,34,36,37,38,39)& 62& 0.19\\
                                                       TW Hya&P(6,7,8,9)& 17& 0.05\\
                            UX Tau A&P(1,2,3,4,5,6,7,8,10,11),R(1,0)& 21& 1.04\\
                                                    VV Ser&P(30,31,32,34,36,37)& 64& 0.72\\
                                                          V1121 Oph&P(31,32,36)&109& 0.04\\
                                                              Wa Oph 6&P(30,38)&139& 0.02\\
\enddata
\tablenotetext{a}{Derived from the velocity at 1.7 $\times$ HWHM and disk inclination.  See the text for details. }
\end{deluxetable*}

\begin{deluxetable}{l|ccc|ccc}
\tablecaption{Model Fits \label{tbl:modelfits}}
\tabletypesize{\scriptsize}
\tablewidth{0pt}
\tablehead{&&$p=-1.5$ &&& $p=0$&\\
 Name &  $R_\mathrm{in}$ &  $R_\mathrm{mid}$ & $q$  &   $R_\mathrm{in}$ & $R_\mathrm{mid}$ & $q$}
\startdata
             AA Tau &      0.10 &      0.20 &       $-3.0$ &     0.010 &      0.20 &      $-3.0$\\
             AB Aur &      0.40 &      0.50 &       $-1.5$ &     0.050 &      1.00 &      $-1.5$\\
           AS 205 N &      0.07 &      0.10 &       $-1.5$ &     0.010 &      0.10 &      $-1.5$\\
             BP Tau &      0.03 &      2.00 &       $-2.0$ &     0.005 &      0.20 &      $-3.0$\\
             DF Tau &      0.10 &      2.00 &       $-2.0$ &     0.020 &      0.50 &      $-2.0$\\
             DL Tau &      0.01 &      1.00 &       $-2.5$ &     0.005 &      0.05 &      $-2.0$\\
            DoAr 44 &      0.20 &      5.00 &       $-3.0$ &     0.005 &      0.50 &      $-2.0$\\
             DO Tau &      0.03 &      0.05 &       $-1.5$ &     0.030 &      0.05 &      $-1.5$\\
             DR Tau &      0.20 &      0.20 &       $-1.5$ &     0.010 &      0.50 &      $-1.5$\\
             GG Tau &      0.05 &      2.00 &       $-2.0$ &     0.005 &      0.06 &      $-1.5$\\
             GK Tau &      0.06 &      2.00 &       $-3.0$ &     0.005 &      0.20 &      $-2.0$\\
             GM Aur &      0.20 &      2.00 &       $-3.0$ &     0.005 &      0.50 &      $-2.0$\\
             GSS 39 &      0.04 &      1.00 &       $-2.0$ &     0.005 &      0.20 &      $-2.0$\\
          HD 135344 B &      1.00 &      1.00 &       $-1.5$ &     0.010 &      5.00 &      $-2.0$\\
          HD 141569 A &      9.00 &     10.00 &       $-3.0$ &     0.100 &     10.00 &      $-1.5$\\
          HD 150193 &      1.00 &      1.00 &       $-3.0$ &     0.200 &      2.00 &      $-3.0$\\
          HD 163296 &      0.30 &      2.00 &       $-3.0$ &     0.005 &      1.00 &      $-3.0$\\
          HD 190073 &      3.00 &      5.00 &       $-1.5$ &     0.005 &      5.00 &      $-1.5$\\
             HL Tau &      0.04 &      5.00 &       $-2.5$ &     0.010 &      0.05 &      $-1.5$\\
           LkHa 330 &      3.00 &      5.00 &       $-1.5$ &     0.100 &     10.00 &      $-1.5$\\
            MWC 480 &      0.05 &      2.00 &       $-2.0$ &     0.005 &      0.20 &      $-2.0$\\
            MWC 758 &      0.20 &      1.00 &       $-2.0$ &     0.050 &      0.50 &      $-2.0$\\
               SR 9 &      0.05 &      2.00 &       $-2.0$ &     0.005 &      0.05 &      $-1.5$\\
              SR 21 &      5.00 &     10.00 &       $-2.0$ &     5.000 &     10.00 &      $-3.0$\\
             SU Aur &      0.20 &      1.00 &       $-3.0$ &     0.020 &      0.50 &      $-3.0$\\
              T Tau &      0.20 &     10.00 &       $-3.0$ &     0.020 &      0.20 &      $-1.5$\\
             TW Hya &      0.10 &      2.00 &       $-3.0$ &     0.005 &      0.50 &      $-3.0$\\
           UX Tau A &      0.30 &      0.50 &       $-1.5$ &     0.005 &      0.50 &      $-1.5$\\
             VV Ser &      1.00 &     10.00 &       $-3.0$ &     2.000 &      2.00 &      $-3.0$\\
          V1121 Oph &      0.04 &      0.50 &       $-2.5$ &     0.005 &      0.10 &      $-2.0$\\
           Wa Oph 6 &      0.04 &      0.20 &       $-3.0$ &     0.005 &      0.10 &      $-3.0$\\
\enddata
\end{deluxetable}

\subsection{Profile Modeling}
\subsubsection{Procedure}
\label{sec:procedure}
While temperature profiles have been measured for outer disk dust \citep[e.g.,][]{Andrews09, Isella09}, the temperature profile
for inner disks or for molecular line-emitting layers is unmeasured.    In fact, it is believed that the disk upper atmosphere is thermally decoupled from the dust, and
its temperature is set by a complex balance of gas heating and cooling \citep{Glassgold01, Kamp04}.   Therefore, we have chosen a model that
makes a minimal number of assumptions about the underlying temperature.  In addition, our model was designed to provide a robust way to determine
the CO inner radius, without the measurement of this radius depending strongly on the choice of model.  In other words, it allows for a good fit to the line wings (and hence $R_\mathrm{in}$) {\it independently} of lower-velocity portions of the line profile.   

We have chosen to model the line profiles as emission from a disk with CO luminosity per unit radius $L_\mathrm{CO}(R)$, where $L_\mathrm{CO, entire\ disk}=\int_{R_\mathrm{in}}^{R_\mathrm{out}} L_\mathrm{CO}(R)dR$ and $L_\mathrm{CO}(R)$ is a broken power law with $R^{p}$ for $R_\mathrm{in}<R<R_\mathrm{mid}$ and $R^{q}$ for $R_\mathrm{mid}<R<R_\mathrm{out}$.  By directly fitting for the shape of $L_\mathrm{CO}(R)$, rather than beginning with the temperature and CO column density profiles, $T(R)$ and $N(R)$, and using these to {\it deduce} $L_\mathrm{CO}(R)$, we make the problem both computationally manageable and simple to interpret.  A similar procedure was used by \citep{Carr04} to model CO and H$_2$O emission from the disk around the young star SVS 13.  

A broken power law is empirically convenient, as it has enough free parameters to properly fit most line profiles (in many cases a single power law is not sufficient to fit the observed line profiles), and yet the inner radius turns out to be well constrained.  A broken power-law approximation for $L_\mathrm{CO}(R)$ also has a physical basis.  At all radii, $L_\mathrm{CO}(R)\sim F(R)\times2\pi R$, where $F(R)$ is the flux emitted by the CO per unit radius.  At small radii, if the gas is optically thick, temperatures are high and so the emission near 5 $\mu$m is in the Rayleigh--Jeans regime (i.e., $F\propto T$).  At larger radii, however, there should be a steeper decrease in $F(R)$ as temperatures drop, the blackbody curve peaks at or beyond 5 $\mu$m, and $F$ drops much more rapidly with $T$.  

In all models, we have assumed $R_\mathrm{out}=100$ AU, as the models are insensitive to $R_\mathrm{out}$ beyond $\sim$ a few AU.  We allow $q$ to take values of $-1.5$, $-2$, $-2.5$, or $-3$.  $q$ primarily affects the degree of double peak in the line profile, but is not well constrained because of degeneracy with other parameters.  We tested values for $p$ from $-2$ to $0$ and, as we will discuss, this parameter has an important effect on the determination of the CO inner radius, $R_\mathrm{in}$.  Values of $R_\mathrm{in}$ from 0.005 to 10 AU were tested, with the lower limit being the approximate radius of a solar mass star and the upper limit producing maximum velocities similar to the NIRSPEC resolution.  We also tested a similar range of values for $R_\mathrm{mid}$, though $R_\mathrm{mid}$ was by definition always larger than or equal to $R_\mathrm{in}$.

 \begin{figure*}
\includegraphics[scale=0.7]{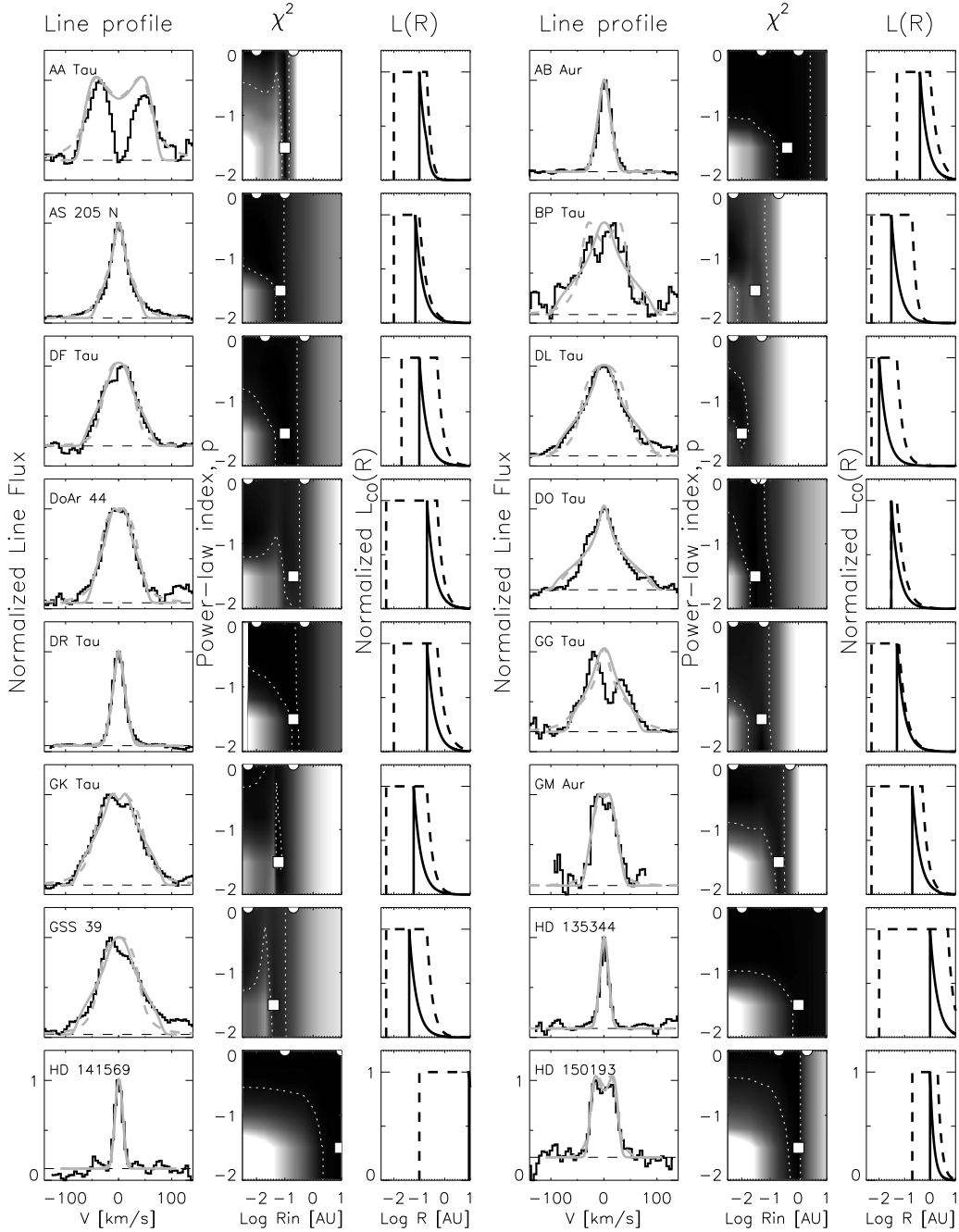}
\caption{Left: composite emission lines and best-fit models with $p=-1.5$.  Middle: $\chi^2$ contours for best models as a function of $R_\mathrm{in}$ and $p$, with dark representing lower $\chi^2$. (Other model parameters are left as free variables). Squares show the best-fit $R_\mathrm{in}$ for $p=-1.5$; semi-circles show the best-fit $R_\mathrm{in}$ and $R_\mathrm{mid}$ for $p=0$.   Dotted lines show 95\% $\chi^2$ confidence intervals, with the reduced $\chi^2$ set to 1 for the best $p=-1.5$ fit.  Right: Contribution to the total line flux as a function of disk radius, $R$  (normalized to 1) for $p=-1.5$ (solid line) and $p=0$ (dotted line).\label{fig:fluxfit_plot1}}
 \end{figure*}
 
 \begin{figure*}
 \addtocounter{figure}{-1}
\includegraphics[scale=0.7]{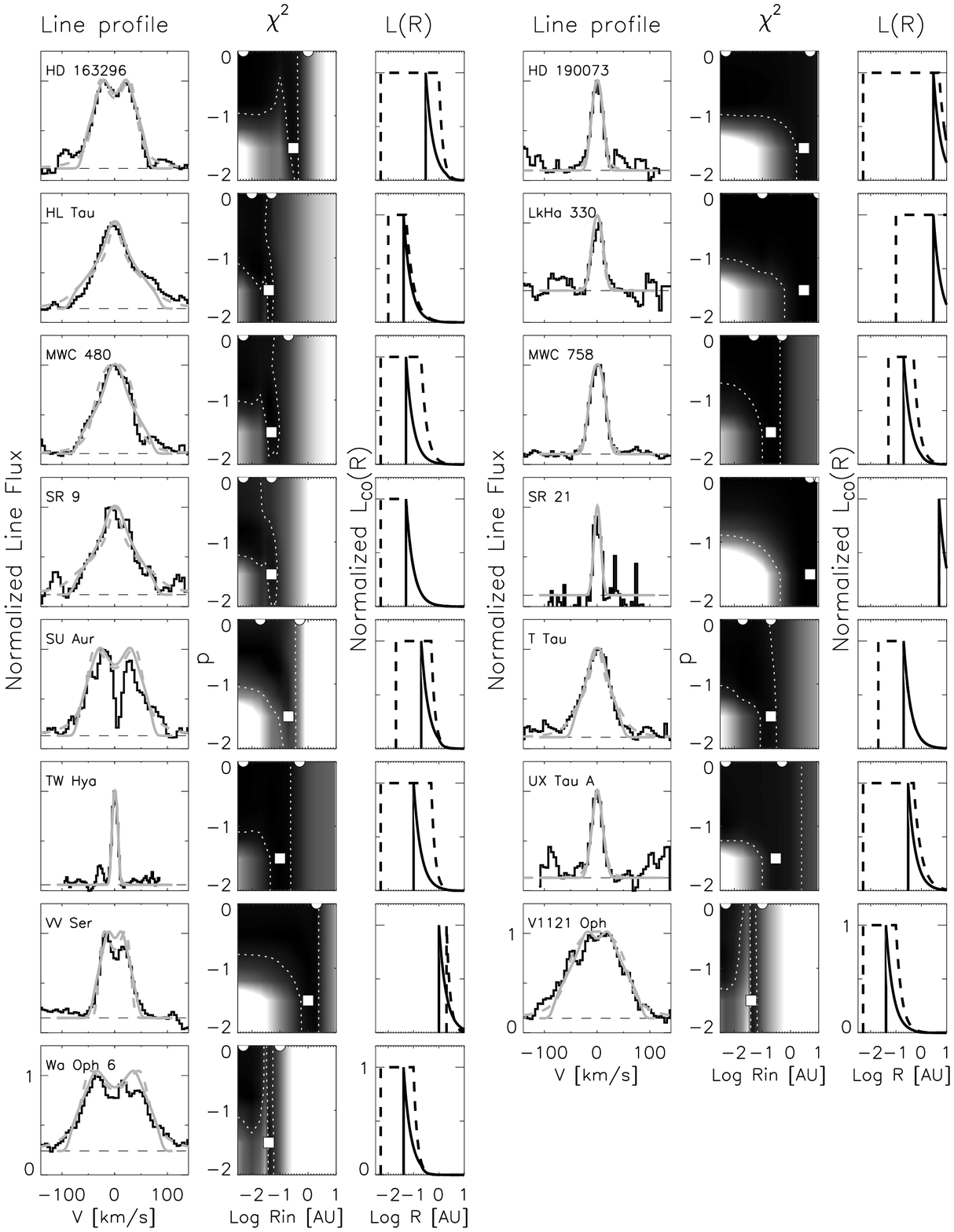}
\caption{{\it Continued}}
 \end{figure*}

\subsubsection{Model Results}
\label{sec:modelresults}
\begin{figure}
\plotone{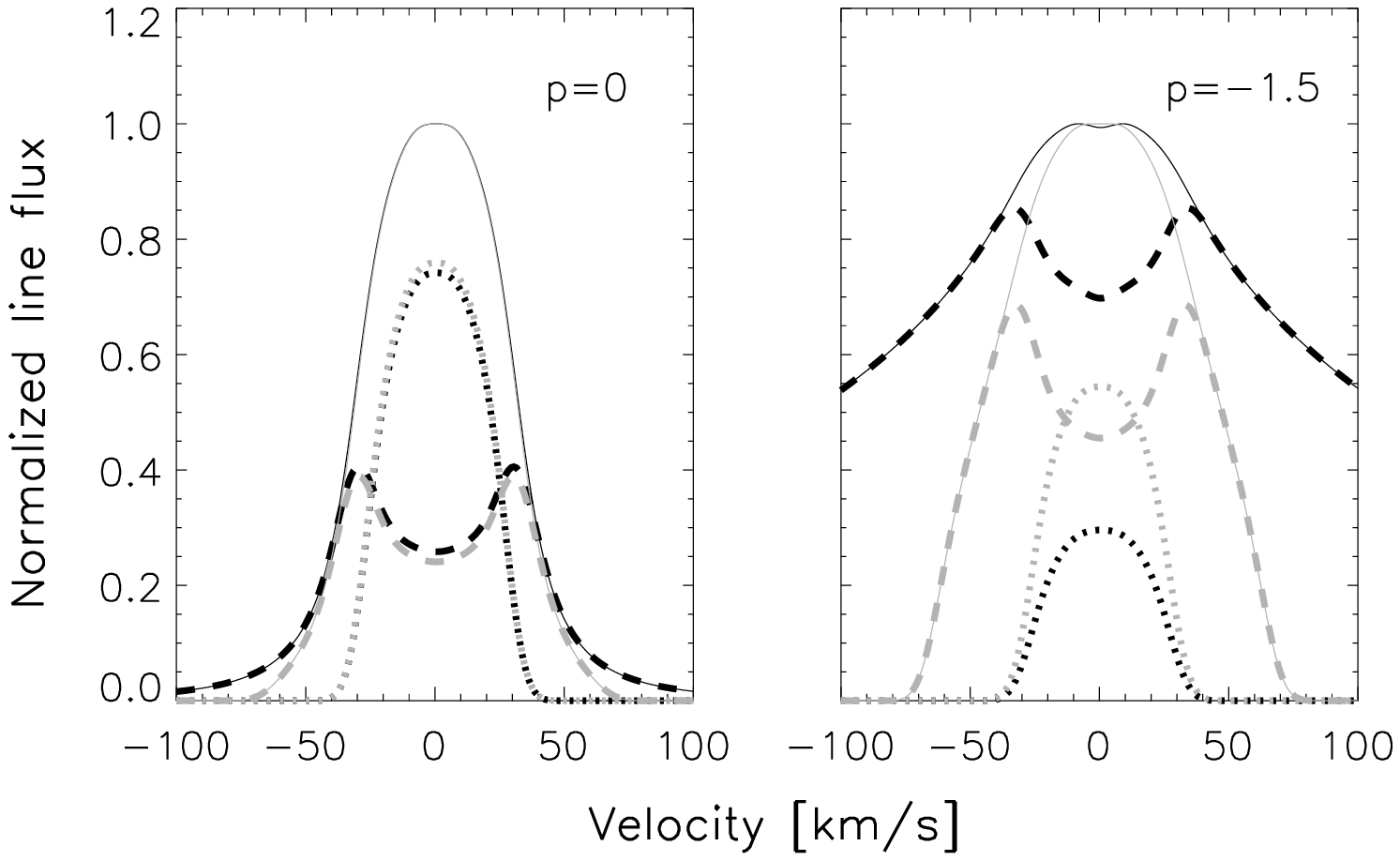}
\caption{Normalized line profiles (solid lines) with $p=0$ (left) and $p=-1.5$ (right) for two different values of $R_\mathrm{in}$: 0.005 AU (black) and 0.1 AU (gray).  All models assume $q=-2$, $R_\mathrm{mid}=0.5$, $R_\mathrm{out}=100$ AU, $M_\star=M_\odot$ and $i=45^\circ$.  Dashed and dotted lines show the contributions from $R_\mathrm{in}$ to $R_\mathrm{mid}$ and $R_\mathrm{mid}$ to $R_\mathrm{out}$, respectively.  Note that with $p=0$ the line profile is relatively insensitive to $R_\mathrm{in}$, while with $p=-1.5$ it is very sensitive to $R_\mathrm{in}$.
 \label{fig:lineplot}}
\end{figure}

Most of the model parameters turn out to be strongly degenerate, so that they cannot be uniquely determined from the line profiles, and since our focus was to study CO inner radii, we did not make an effort to understand these parameters in detail.   Instead, our goal was to investigate whether 
$R_\mathrm{in}$ could be determined robustly.  We find that if $p$ is strongly negative, then the line shape becomes very sensitive to $R_\mathrm{in}$.  This is demonstrated in the middle panels of Figure \ref{fig:fluxfit_plot1}, which show $\chi^2$ as a function of $R_\mathrm{in}$ and $p$, with all other parameters being free variables.  If $p\lesssim -1.5$, then $R_\mathrm{in}$ is robustly determined.  If $p$ is 0 (i.e., $L_\mathrm{CO}(R)$ is flat from $R_\mathrm{in}$ to $R_\mathrm{mid}$), then the line shape becomes insensitive to $R_\mathrm{in}$, but is instead sensitive to $R_\mathrm{mid}$.  (For intermediate values of $p$, the line shape is sensitive to a combination of $R_\mathrm{in}$ and $R_\mathrm{mid}$.)  Thus, the line profile is sensitive to steep drops in $L_\mathrm{CO}(R)$.  This is due to the fact that in order to be sensitive to some $R_0$ (either $R_\mathrm{in}$ or $R_\mathrm{mid}$), any change to this parameter, $\Delta R_0$, must result in a significant change to the total line luminosity, i.e., $L(R_0)\Delta R_0$ must be a significant fraction of $\int_{R_\mathrm{in}}^{R_\mathrm{out}} L_\mathrm{CO}(R) dR$.    In Figure \ref{fig:lineplot}, we show line profiles with $p=0$ and $p=-1.5$ for two different values of $R_\mathrm{in}$, and demonstrate that with $p=-1.5$ the line profile is very sensitive to $R_\mathrm{in}$, but with $p=0$, the line profile changes only slightly with changes in $R_\mathrm{in}$.   

Due to the dependence of $R_\mathrm{in}$ and $R_\mathrm{mid}$ on $p$, we have explored two classes of solutions that can fit the data well in most cases.  In the first approach, $p$ is set to $-1.5$, and $R_\mathrm{in}$ is well-determined.  In the second $p$ is set to $0$, and $R_\mathrm{mid}$ is well-determined.  The best-fit parameters in these two cases are presented in Table \ref{tbl:modelfits}.  The best-fit line profiles and corresponding $L_\mathrm{CO}(R)$ profiles are shown in Figure \ref{fig:fluxfit_plot1}.  The $p=0$ case places no constraints on the inner edge of the CO emission and has a sudden steep drop-off in $L_\mathrm{CO}(R)$ occurring at $R_\mathrm{mid}$.  The $p=-1.5$ case has a distinct region free of CO between $R_\star$ and $R_\mathrm{in}$.   We believe that the second class of models is more physically motivated, as CO can be cleared out near the star due to photodissociation or disk truncation.  However, we cannot empirically rule out the other class of models.  There are a few sources in which the $p=0$ solution results in a significantly-improved fit --- most notably AS 205 N and Wa Oph 6 (see Figure \ref{fig:fluxfit_plot1}).  However, AS 205 N is known to be a member of a small subset of disks with ``peaky'' line profiles \citep{Bast11}, in which low-velocity flux may be enhanced by a low-velocity disk wind \citep{Pontoppidan11}. Thus, a Keplerian disk model is probably not appropriate in this case, and we do not believe that the model with $p=0$ is necessarily more physically realistic than the model with $p=-1.5$.  

\begin{figure}
\plotone{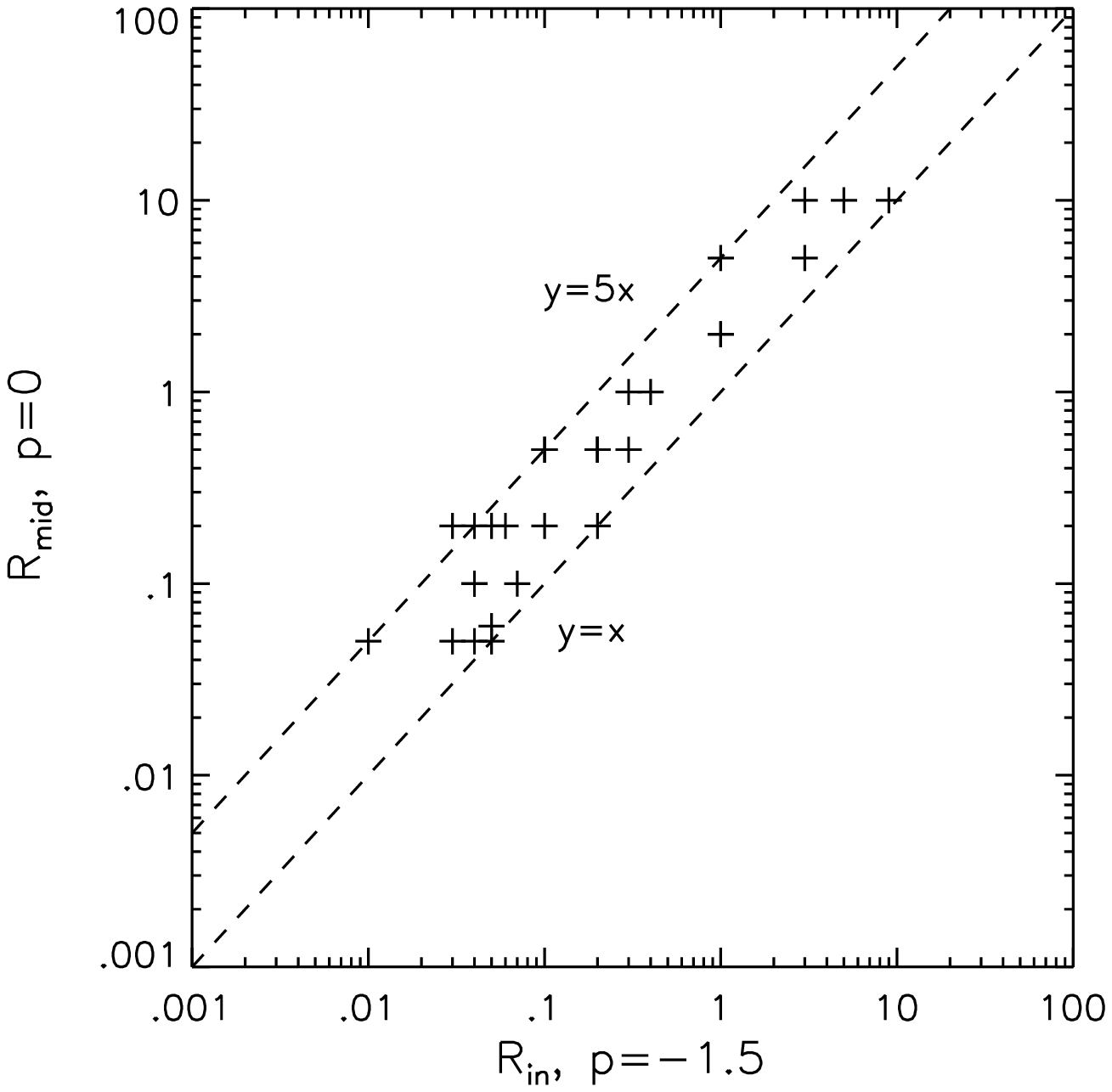}
\caption{Plot of $R_\mathrm{in}$ with $p$ fixed at $-1.5$ and $R_\mathrm{mid}$ with $p$ fixed at $0$.  $R_\mathrm{mid}$ with $p=0$ is typically $1-5\times R_\mathrm{in}$ with $p=-1.5$ (see dashed lines). \label{fig:rin_rmid}
}
\end{figure}

In most cases, one class of fits is not preferred over the other, and since we feel the $p=-1.5$ case is more physically motivated, we adopt this model for the remainder of this work.   However, in Figure \ref{fig:rin_rmid}, we show that $R_\mathrm{in}$ with $p=-1.5$ and $R_\mathrm{mid}$ with $p=0$ are related such that $R_\mathrm{mid}\sim1-5\times R_\mathrm{in}$ (because with $p=0$ there is still some flux contribution from the region inside $R_\mathrm{mid}$). Therefore, uncertainty about the choice of model simply means that there is a systematic uncertainty in the location of the drop-off in $L_\mathrm{CO}$ by a factor of a few, and any {\it trends} in $R_\mathrm{in}$ should be robust.

When comparing $R_\mathrm{in}$ with other disk radii elsewhere in the paper, we will refer to our best-fit $R_\mathrm{in}$ as $R_\mathrm{CO}$.

\subsubsection{Error Estimates}
\label{sec:errors}
Assuming zero error in disk inclination, it is apparent from the $\chi^2$ diagrams in Figure \ref{fig:fluxfit_plot1} that $R_\mathrm{in}$ is uncertain to $\sim$ 0.5 dex, or factors of a few.  Uncertainties are also introduced due to errors in disk inclination.  These scale roughly as $\Delta R/R \sim \cos{i} \Delta i$, and so are largest for low-inclination disks.  With a typical inclination ($45^\circ$) and error ($\pm10^\circ$), errors in $R$ (derived from Kepler's law) are $+50\%$ and $-25\%$.  Thus, they are similar to or somewhat smaller than the systematic uncertainties.  However, for smaller inclinations, uncertainties in $i$ may dominate.  In any case, an uncertainty of $\sim$ 0.5 dex should reasonably be assumed for $R_\mathrm{in}$.  

For a few sources with narrow line profiles, it is apparent from the lack of a large radius cutoff in $\chi^2$ (Figure \ref{fig:fluxfit_plot1}) that the models are fairly insensitive to differences in $R_\mathrm{in}$ at radii larger than the best-fit value.  These include AB Aur, HD 135344 B, HD 141569 A, HD 190073, LkH$\alpha$ 330 and SR 21.

As discussed in Section \ref{sec:modelresults}, there is a systematic uncertainty introduced by the two classes of models capable of reproducing the observed line shapes.  
There is also a potential bias introduced by the choice of lines used in the composite line profile.  In theory, one should be able to obtain the same $R_\mathrm{in}$ from all emission lines and we have tested whether using only low-excitation lines ($J\lesssim12$) yields the same result.  In practice, we find that utilizing low-excitation lines only, the derived $R_\mathrm{in}$ are similar to or larger than the nominal $R_\mathrm{in}$, with the ratio typically being 1--2.  Although biased towards larger $R_\mathrm{in}$, this is within our known uncertainty of a factor of a few, and we believe that the high-excitation lines are likely to be more sensitive to $R_\mathrm{in}$, and so yield the more accurate fit.

\subsubsection{Benchmarking to Other Work}
\label{sec:benchmark}
At least two other methods for determining $R_\mathrm{in}$ have appeared in the literature.  In one, a CO temperature (and density) profile is assumed, and $R_\mathrm{in}$ is fit as a single free variable in a disk model \citep[e.g.,][]{Blake04, Salyk09, Bast11}.  In our model, in contrast, the temperature, density, level populations, and emitting area are all wrapped up into $L_\mathrm{CO}(R)$.  To compare the two models, we fit model line profiles, constructed using the procedure described in \citet{Salyk09}, with the simple model described here, setting $p=q$.  We find that disk temperature profiles of the form $T\propto R^{\rho}$ with $\rho=-0.6$ are equivalent to
models with $L_\mathrm{CO}(R)\propto R^{-3}$, and $T\propto R^{-0.2}$ is equivalent to $\sim L_\mathrm{CO}(R) \propto R^{-1.5}$. 

Compared to the values derived by \citet{Salyk09}, we find similar radii for many sources, with the notable exception of HD 141569 A, LkH$\alpha$ 330, and SR 21, for which 
\citet{Salyk09} find radii larger by factors of a few to $\sim$ 10.  All three of these sources, however, are ones in which $\chi^2$ is not sensitive to $R_\mathrm{in}$ 
at radii larger than the best-fit value.   We also find radii consistent with \citet{Bast11} for AS 205 A and TW Hya, who estimate inner radii of 0.04 and 0.1 AU, respectively (although they caution that the parameter space for these models was not well explored; J.~E.\ Bast, private communication).  Our respective results are {\it inconsistent} for VV Ser, for which Bast et al.\ derive an inner radius of 0.08 AU and we derive an inner radius of 0.72 AU; however, a closer look at their model reveals that they are in a regime similar to our $p=0$ case, in which $\chi^2$ is simply not very sensitive to the choice of inner radius.

\begin{figure}
\plotone{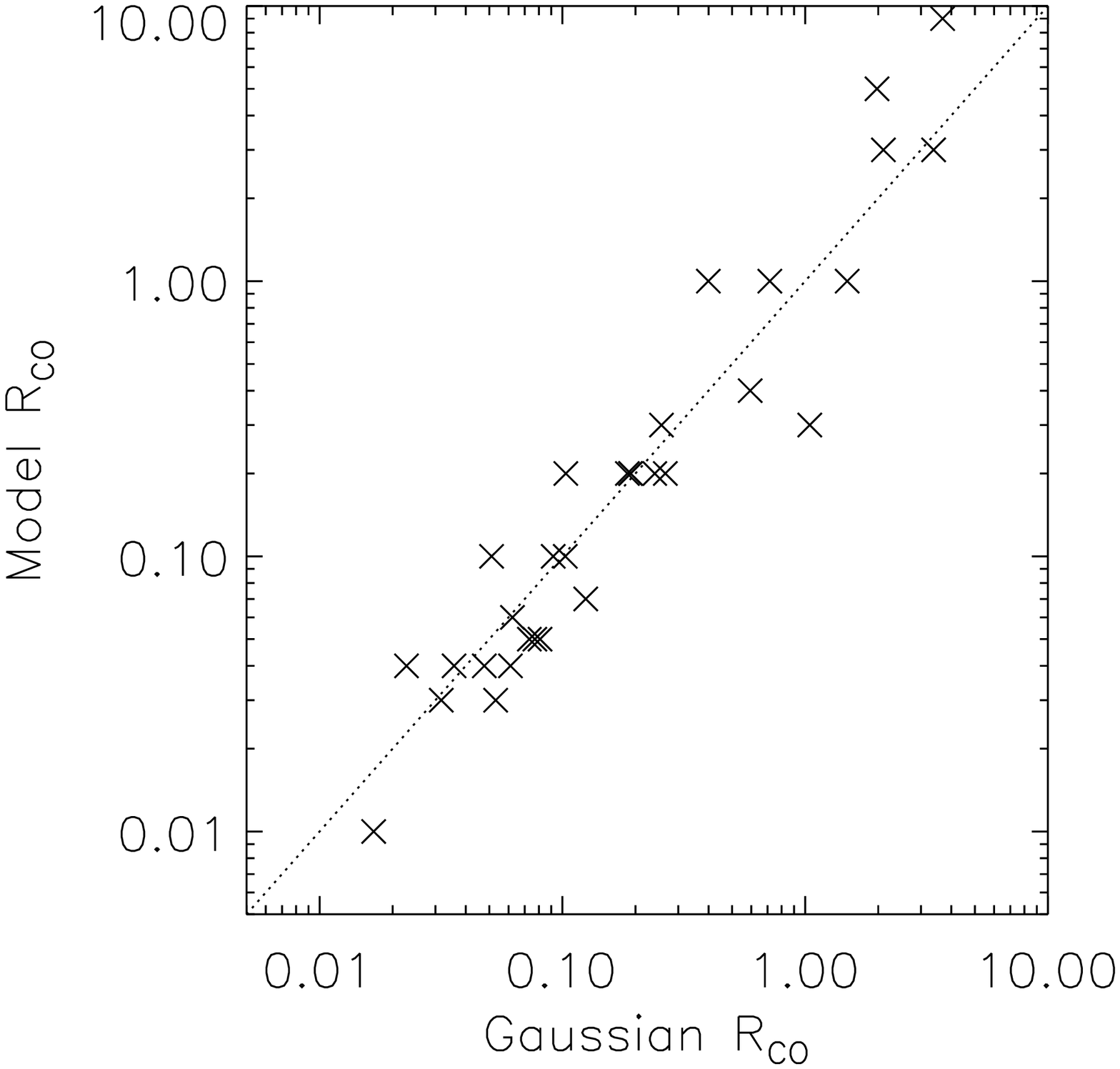}
\caption{Modeled inner radii compared with inner radii derived from Gaussian fits and velocities at 1.7$\times$HWHM (with the factor of 1.7 determined empirically). \label{fig:radcomp}}
 \end{figure}

Another common approach to estimating $R_\mathrm{in}$ is to simply choose some velocity (typically either 2$\times$ the half-width at half maximum (HWHM), or the half-width at zero intensity),
and set $R_\mathrm{in}$ to the radius with that Keplerian velocity.  The difficulty of this approach is that it is not obvious which velocity to choose.  In Figure \ref{fig:radcomp}, we compare
$R_\mathrm{in}$ to inner radii derived from fitting the profiles with single or double (emission plus absorption) Gaussians ($R_\mathrm{gauss}$).  In particular, we show solutions
in which $R_\mathrm{in}$ is derived from the velocity at 1.7$\times$ HWHM, which comes closest to reproducing our results.   Deviations from 1:1 are of order a factor of a few, and
so this simple approach is remarkably consistent with our more complex model.  Thus, we suggest that using the velocity at 1.7$\times$ HWHM is a reasonable choice for calculating $R_\mathrm{in}$ using simple Gaussian fits.  We use this result to incorporate LkCa 15 into our analysis, utilizing the FWHM measured by \citet{Najita03}.
This result also gives us confidence that the model-derived $R_\mathrm{in}$ is reflecting the bulk line shape and FWHM, and is not instead some spurious result
heavily biased by the line/continuum ratio, the noise level, or any other aspects of the data.

\subsubsection{Comparison with Spectro-astrometric Results}
\label{sec:sa}
Four sources in our sample (HD 135344 B, SR 21, TW Hya, and VV Ser) have inner radii derived from a combined line shape and spectro-astrometric (SA) profile analysis \citep{Pontoppidan08,Pontoppidan11}.  Although the overlapping sample is small, and two of these are sources for which we do not have a strong upper limit on $R_\mathrm{in}$ (see Section \ref{sec:benchmark}), our results appear broadly consistent with these results.  \citet{Pontoppidan08} measured CO inner radii for HD 135344 B , SR 21 and TW Hya assuming a power-law disk temperature profile, and derived values within a factor of 3.5 of our $R_\mathrm{in}$.
Note that the SA signal is most sensitive to the {\it extent} of the CO emission, and also provides a hard upper limit on $R_\mathrm{in}$.  However, the SA profile is
not very sensitive to $R_\mathrm{in}$, and so modeling that incorporates SA is subject to the same uncertainties in the gas temperature profile, and does not necessarily
determine $R_\mathrm{in}$ more accurately.  


\citet{Pontoppidan11} report SA radii, which they define as the radius at the peak of the SA profile, for HD 135344 B, SR 21, TW Hya and VV Ser.   The SA signal is the flux-weighted mean position of the emission at each velocity and is therefore sensitive to the distribution of emission rather than simply the inner boundary.  The SA radii will therefore always be larger than $R_\mathrm{in}$ unless the emitting region is infinitesimally thin.  
In the context of our two-power-law model, the SA radii would be affected by $p$, $q$ and $R_\mathrm{mid}$, since these all affect the amount of flux at large radii.  Thus, the SA profiles are complementary to the results from line shape analysis, and the two can potentially be used in concert to derive the shape of $L_\mathrm{CO}(R)$.  The SA radii ($R_\mathrm{SA}$)  derived by \citep{Pontoppidan11} are factors of 2--5 larger than our $R_\mathrm{in}$.  We have tested our two-power-law models with a simple code to calculate $R_\mathrm{SA}$ and find that we can simultaneously reproduce the observed $R_\mathrm{in}$ and $R_\mathrm{SA}$ by adjusting other model parameters.  Therefore, our $R_\mathrm{in}$ are consistent with and complementary to the SA results.


\section{Line Flux Models \label{sec:lines}}
Rotation diagrams can be used to derive characteristic column densities, emitting areas and temperatures for the CO emitting layer.  Rotational levels are assumed to be populated according to LTE.  Although these parameters are not fully realistic, since the emission actually comes from a range of radii and heights in the disk atmosphere, they provide a convenient way to roughly characterize and compare the emission within a sample of disks.  In this work, we will also compare these parameters with $R_\mathrm{CO}$.

\subsection{Fitting Procedure}
We fit CO line fluxes with an LTE slab model following the procedure described in detail in \citet{Salyk09}.   Sources with fewer than two measurable line fluxes with $J_\mathrm{up}>20$ or
a significant amount of self-absorption in low-excitation lines were excluded from the analysis, as these did not provide reliable model fits.  Line fluxes were calculated using Gaussian fits to the emission lines, and rotation
diagrams were fit using a grid of LTE slab models with a single temperature ($T$), CO column density ($N$), and emitting area ($A$) as free parameters.   In addition, $^{13}$CO detections or non-detections were used to constrain $N$ and remove model degeneracies.   Best-fit model parameters are listed in Table \ref{tbl:rot}. 

\begin{deluxetable}{cccc}
\tabletypesize{\scriptsize}
\tablecaption{Rotation Diagram Fit Results \label{tbl:rot}}
\tablehead{\colhead{Star}&\colhead{$T$ (K)\tablenotemark{a}}  &  \colhead{log($N$ (cm$^{-2}$))}  &  \colhead{log($A$ (AU$^2$))}} 
\startdata
AA Tau& 950& 18.6& $-1.3$\\
AB Aur& 600& 18.8& $0.3$\\
AS 205 N& 975& 18.7& $-0.3$\\
DF Tau&1025& 19.1& $-0.7$\\
DL Tau&1675& 18.6& $-1.7$\\
DoAr 44&1150& 18.2& $-1.6$\\
DO Tau&1575& 19.0& $-1.7$\\
DR Tau&1250& 18.9& $-1.0$\\
GG Tau&1500& 18.6& $-2.1$\\
GK Tau&1600& 18.1& $-1.5$\\
GSS 39&1675& 18.1& $-1.4$\\
HD 135344 B& 900& 17.9& $-1.5$\\
HD 141569 A& 275& 18.3&$$2.0$$\\
HD 150193& 700& 18.1& $-0.5$\\
HD 163296& 825& 18.5& $-0.5$\\
LkHa 330& 850& 17.4& $-0.6$\\
MWC 480& 975& 18.0& $-0.8$\\
MWC 758& 700& 18.3& $-0.2$\\
SR 9&1575& 16.7& $-0.8$\\
T Tau&1325& 17.9& $-0.6$\\
TW Hya& 700& 17.7& $-1.9$\\
V1121 Oph&1125& 18.5& $-1.4$\\
Wa Oph 6&1675& 18.3& $-1.6$\\
\enddata
\tablenotetext{a}{It is not straightforward to define error bars, because of the degeneracy of model parameters \citep{Salyk09}.  Typical uncertainties are $\pm200$ K for $T$, and $\pm$0.5 for $\log(N)$ and $\log(A)$.}
\end{deluxetable}

\section{Discussion}
\subsection{Correlation with Corotation Radii}
\begin{figure}
\plotone{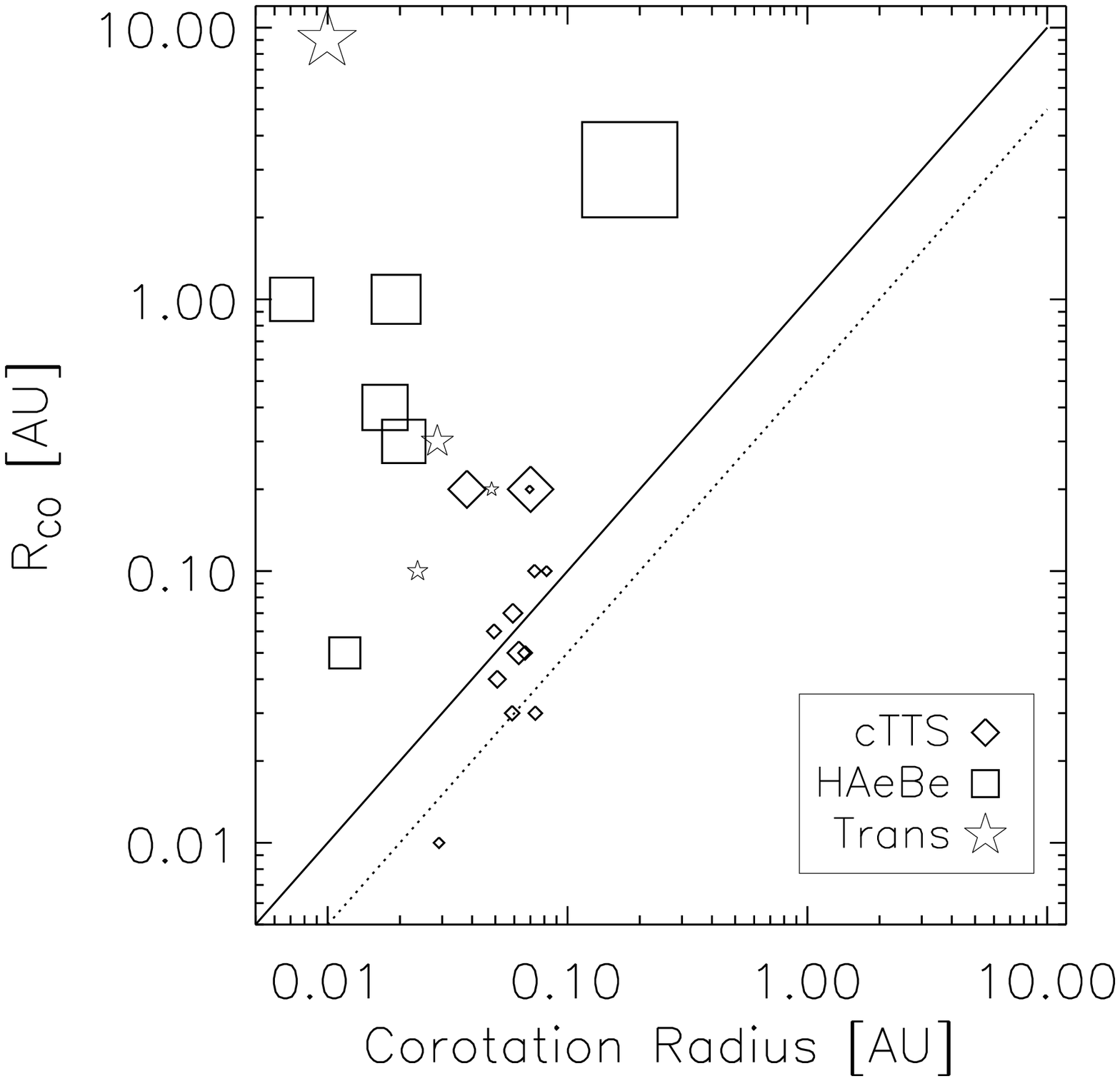}
\caption[CO inner radius against corotation radius]{CO inner radius against corotation radius.  The size of the symbol is proportional to the stellar mass.  The solid line shows a 1:1 correspondence, while the dotted line has $R_\mathrm{CO}=0.5\times R_\mathrm{corot}$. \label{fig:corot} }
 \end{figure}

For a sample of five T Tauri stars observed by \citet{Najita03}, CO inner radii were found to be in the range $\sim0.5-1$ times the corotation radius, the radius
at which the disk's Keplerian angular velocity equals that of the stellar surface:
\begin{equation}
R_c=\frac{GM_\star R_\star^2 \sin^2(i)}{(v\mathrm{sin}(i))^2}.\label{eq:corot}
\end{equation}
For convenience, we assume that $i_\star=i_\mathrm{disk}$, although the two need not be exactly the same.
$R_c$ is expected to be similar to the magnetic truncation radius, within which the disk is cleared by magnetospheric accretion \citep{Shu94}.  In Figure \ref{fig:corot},
we plot $R_\mathrm{CO}$ against $R_c$, as well as lines representing $R_\mathrm{CO}=R_c$ and $R_\mathrm{CO}=0.5\times R_c$.  Symbol sizes are proportional to
the stellar mass.  We find that $R_\mathrm{CO}$ is consistent with being $\sim$ 0.5--1$\times$ the corotation radius  {\it for T Tauri disks}, but that this relationship 
quickly breaks down for transitional disks and disks around HAeBe stars.  Additionally, there is a hint that the degree of discrepancy scales with stellar mass, a trend we discuss in Section \ref{section:size-luminosity}.

\subsection{A Size-Luminosity Relationship for Gas Disks}
\label{section:size-luminosity}
There exists a known correlation between inner dust disk size and stellar luminosity, in which inner dust disk sizes appear to be determined by dust sublimation, with
$T_\mathrm{sub}\sim1000-1500 \mathrm{K}$ \citep[see][and references therein]{Dullemond10}.  For blackbody grains in a disk with an optically thin inner region,
\begin{equation}
T_\mathrm{sub}\approx \left(\frac{L_\star}{16\sigma\pi R_\mathrm{sub}^2}\right)^{1/4}.\label{eq:tsub}
\end{equation}
This formula neglects many complications, including grain properties, the pressure dependence of sublimation temperature, and grain scattering, which are discussed
in detail in \citet{Dullemond10}; since the relative importance of these various effects has not yet been resolved, we work with this simple version.
 
\begin{figure*}
\plotone{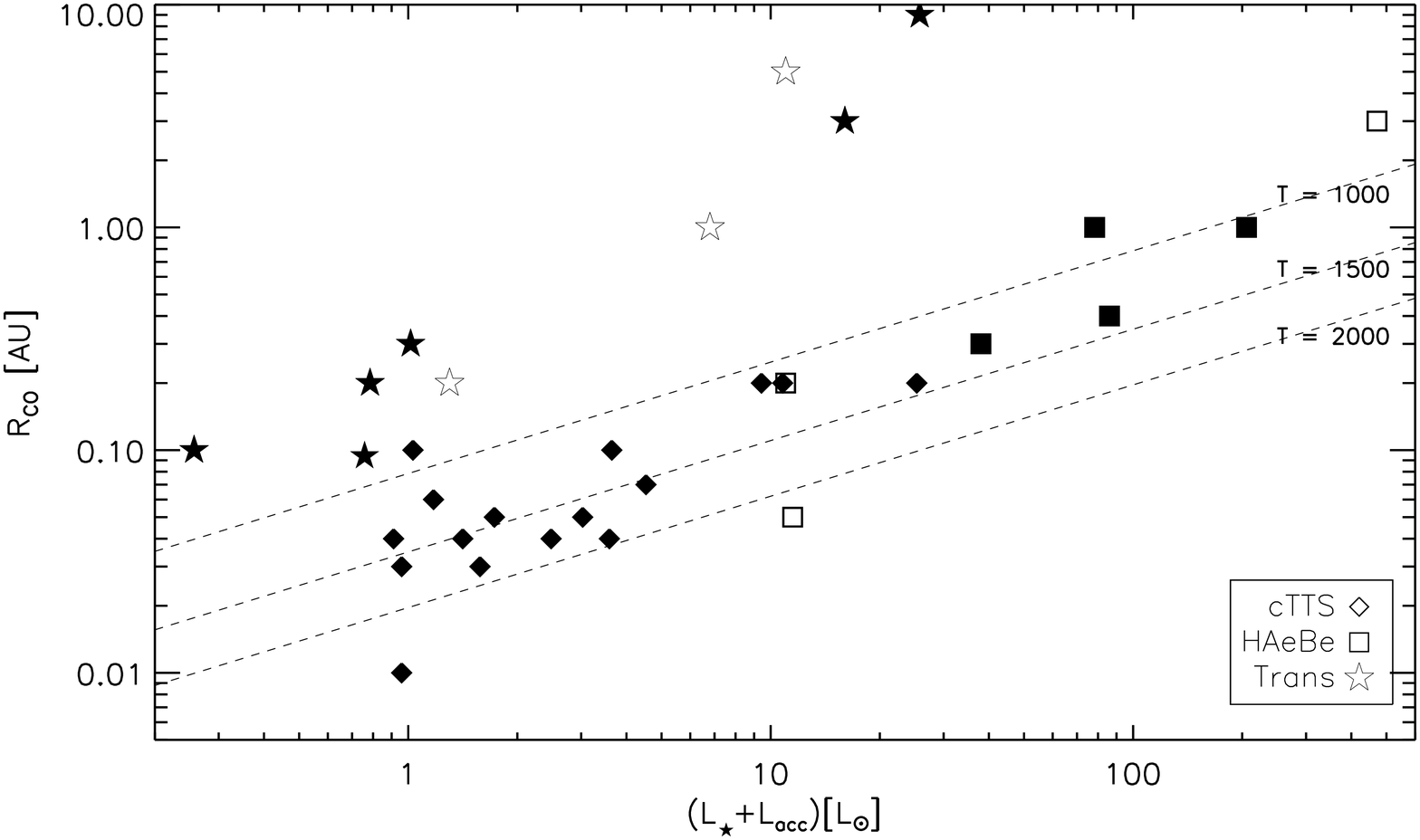}
\caption[CO inner radius against luminosity]{CO inner radius against luminosity. Diamonds are cTTs disks, squares are HAeBe disks and stars are transitional disks.   Filled symbols have measured accretion rates, which are incorporated into $L_\mathrm{acc}$, while unfilled symbols have $L_\mathrm{acc}=0$.  Dashed lines show theoretical curves for dust sublimation radius as a function of luminosity.  \label{fig:lrin_acc}}
 \end{figure*}

\begin{figure}
\plotone{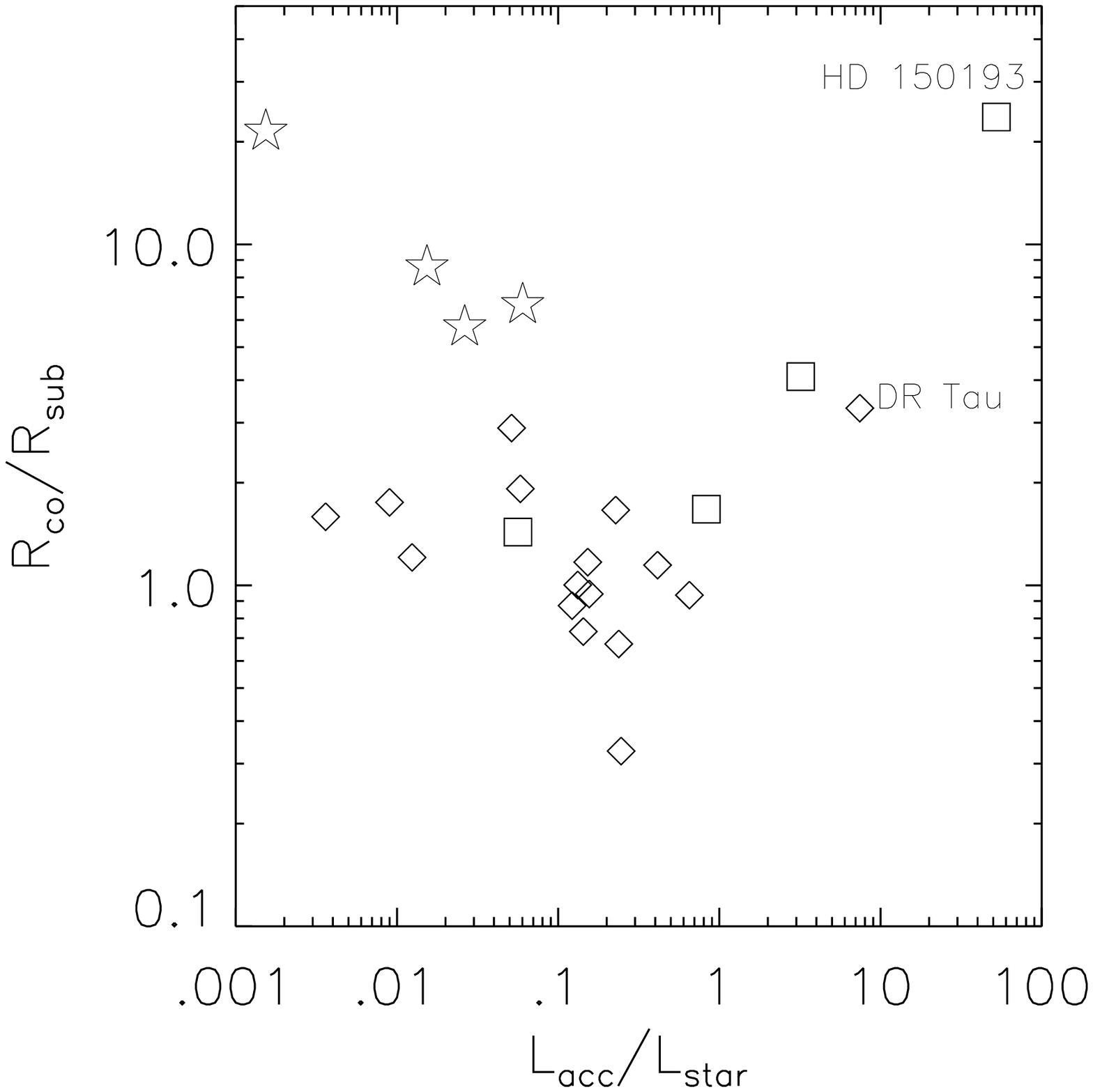}
\caption{$R_\mathrm{CO}/R_\mathrm{sub}$ against $L_\mathrm{acc}/L_\star$, with $R_\mathrm{sub}$ calculated from $L_\star$ alone, and assuming $T_\mathrm{sub}=1500$ K.  Symbols are the same as in Figure \ref{fig:lrin_acc}. \label{fig:acc}}
\end{figure}

CO inner radii might also be set by dust sublimation, since CO can be dissociated by UV radiation if grains are not available for shielding.  However, previous
studies of CO inner radii were limited to small ranges in stellar luminosity  \citep{Najita03, Blake04, Salyk09}.  In addition, low-mass cTTs's have dust sublimation radii
that are similar in size to the stellar corotation radii, so distinguishing between the two effects is difficult.  In Figure \ref{fig:lrin_acc}, we show the relationship between luminosity and $R_\mathrm{CO}$ for our complete sample of disks.  Note that instead of simply plotting the stellar luminosity, we also include the accretion luminosity if the accretion rate is known, as this is the total luminosity seen by the disk.  We assume $L_\mathrm{acc}=0.8GM_\star \dot{M}/R_\star$ \citep{Hartmann98}.  Neglecting to include the accretion contribution to the luminosity yields outliers with high accretion rates (see Figure \ref{fig:acc}).

In Figure \ref{fig:lrin_acc} we show the strong correlation between $R_\mathrm{CO}$ and $L_\star+L_\mathrm{acc}$, consistent with
dust sublimation temperatures of $T\sim1500-2000$ for cTTs disks and slightly lower temperatures, $\sim1000-1500$ for HAeBe disks.  Transitional disk radii are larger, but also follow a similar trend (discussed further in Section \ref{sec:trans}).  Including all disks, the $p$-value associated with linear regression of $\log(R_\mathrm{CO})$ against $\log(L_\star+L_\mathrm{acc})$ is $6\times 10^{-5}$, and excluding transitional disks, it is $10^{-10}$, so the increase in radius with luminosity is highly statistically significant. Therefore, we believe this is strong evidence for a dependence of $R_\mathrm{CO}$ on dust sublimation.

The slope of the trend (excluding transitional disks), $0.7 \pm 0.1$, is somewhat steeper than the slope of $0.5$ expected from dust sublimation alone, at the 2 $\sigma$ level.  In more detail, $R_\mathrm{CO}$ may be set by a balance of photodissociation (which eats outward in the disk) and accretion (which replenishes inward).  This could result in relatively smaller $R_\mathrm{CO}$ for T Tauri disks and larger for HAeBe disks due to the respectively lower and higher photodissociating UV fluxes.

\subsection{Comparison with Interferometric Dust Inner Radii}
\begin{figure}
\plotone{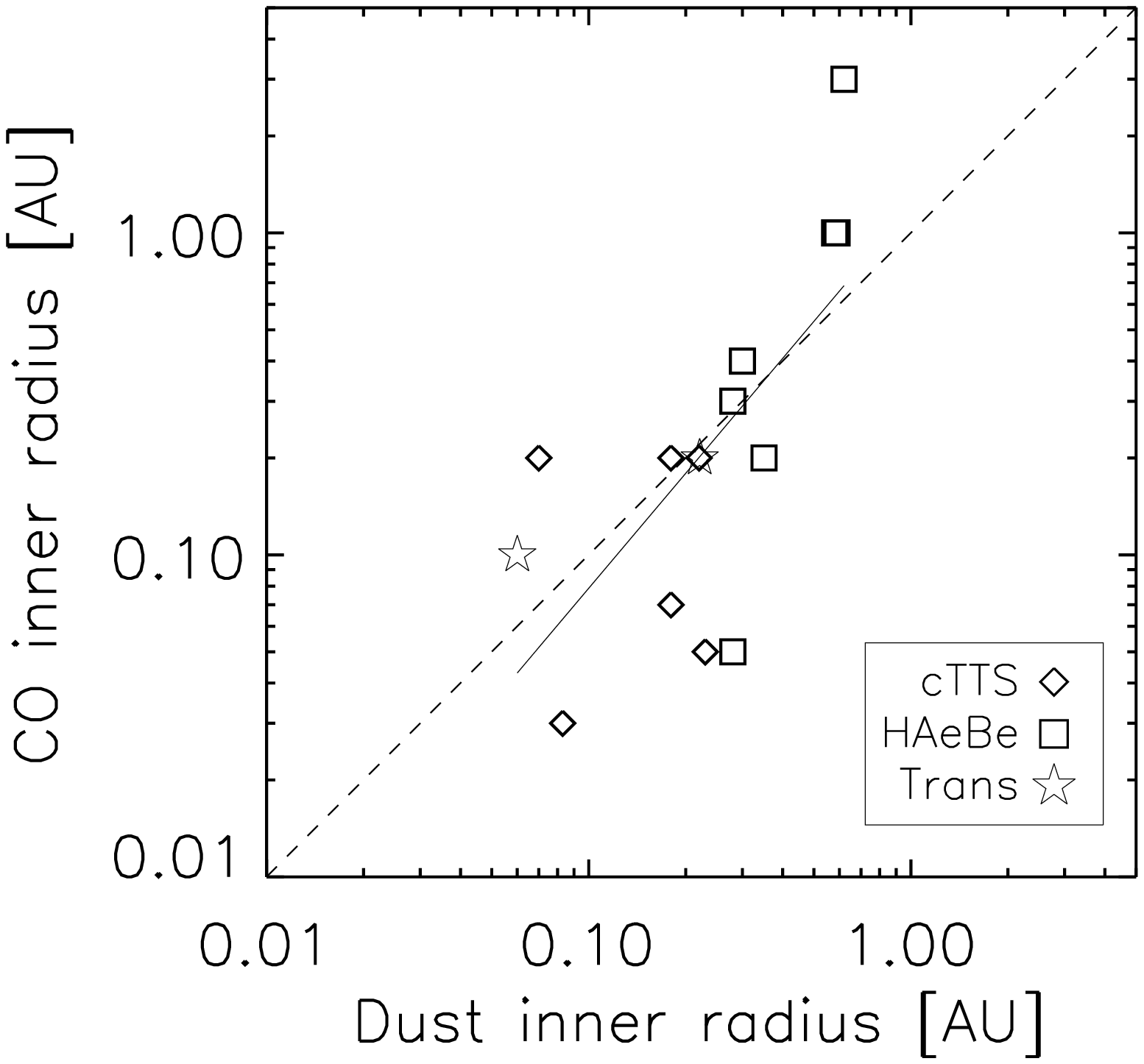}
\caption[CO inner radius against dust inner radius]{CO inner radius against dust inner radius. Symbols are the same as in Figure \ref{fig:lrin_acc}.  The solid lines shows the best linear fit, and the dashed line shows a 1:1 correspondence. \label{fig:gas_dust}}
\end{figure}

If dust acts as a shield against photodissociation, then we should also expect a correlation between $R_\mathrm{CO}$ and measured dust inner radii from near-IR interferometers.  We show these radii in Figure \ref{fig:gas_dust}, along with a line marking a 1:1 correlation, and another marking the best linear fit.  Although there is a statistically significant correlation between the two variables, there is also significant scatter, of order 0.5 dex.  Uncertainties in $R_\mathrm{CO}$ can be a factor of a few (see Section \ref{sec:errors}), which may account for some of the scatter.  Also, dust inner radii are measured using different models in different studies (including thin rings, disks, both inclined and not inclined), and so it is possible that some of the scatter arises from the choice of dust model, or from dust models not accounting for the disk inclination.  Using a coherent sample of interferometric visibilities, analyzed in the same way, and accounting for disk inclination, could test this hypothesis.  Another possibility is that there is real scatter due to different rates of photodissociation and/or replenishment via accretion.  This is consistent with the observation that several HAeBe disks have anomalously large $R_\mathrm{CO}$, while several cTTs's have anomalously small $R_\mathrm{CO}$.

\subsection{Transitional Disks}
\label{sec:trans}
\begin{figure}
\plotone{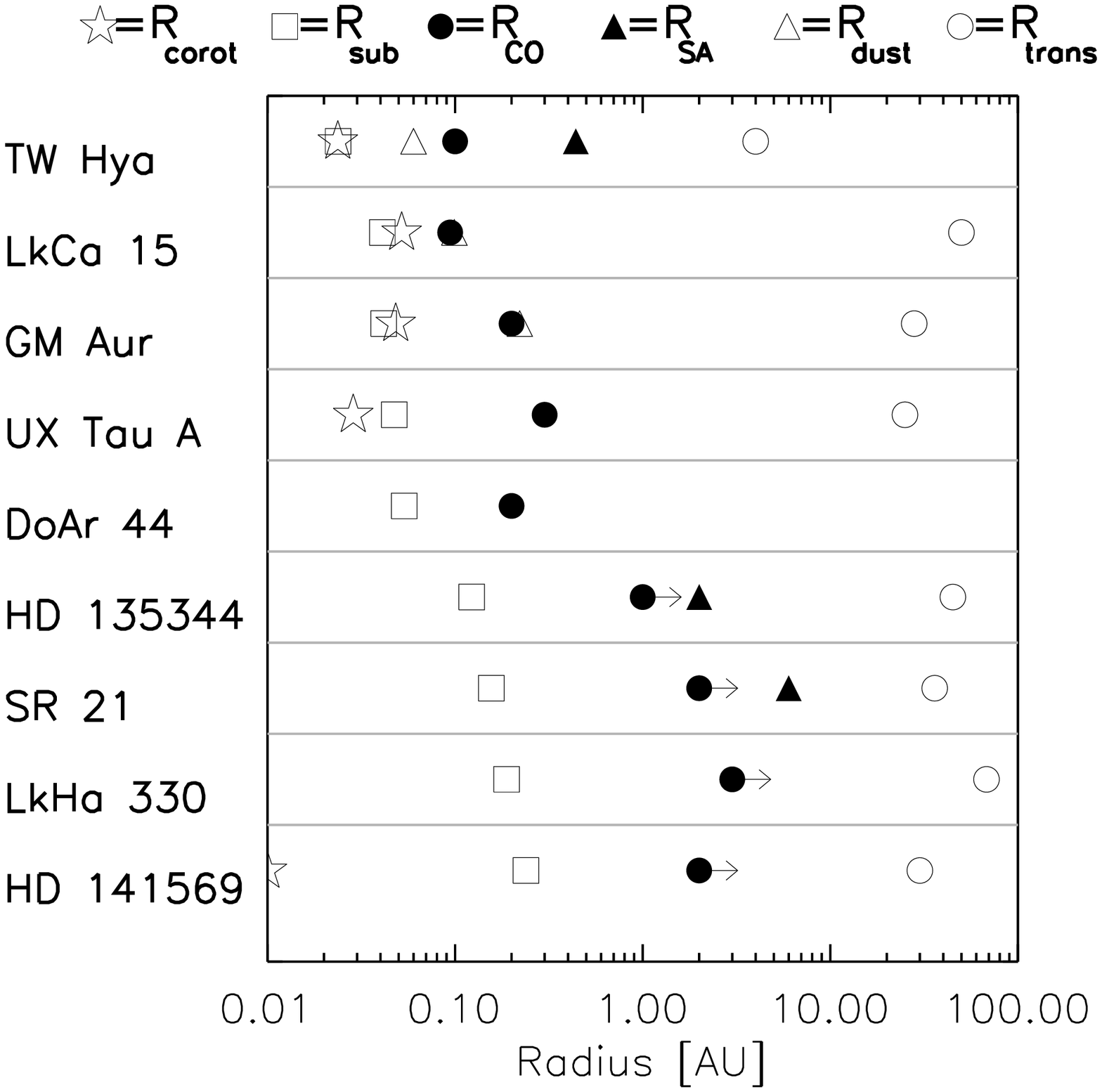}
\caption{Comparison between $R_\mathrm{in}$, $R_\mathrm{sub}$, $R_\mathrm{CO}$, $R_\mathrm{SA}$, $R_\mathrm{trans}$, and other system radii, for transitional disks. ($R_\mathrm{SA}$ from \citealt{Pontoppidan11}; $R_\mathrm{trans}$ from \citealt{Marsh02}, \citealt{Brown09}, \citealt{Akeson11}, and \citealt{Andrews11}.  Other parameters from Table \ref{tbl:params}.)
\label{fig:radplot_trans}}
\end{figure}

As discussed in several prior studies \citep{Najita03,Rettig04,Salyk09}, $R_\mathrm{CO}$ lies well inside $R_\mathrm{trans}$ for many transitional disks (where $R_\mathrm{trans}$ is the radius at which the disk becomes optically thick).  This result has also been confirmed via spectroastrometry for HD 135344 B and TW Hya \citep{Pontoppidan08}.  In Figure \ref{fig:radplot_trans}, we closely reproduce the results of \citet{Salyk09}, who showed that
the best-fit $R_\mathrm{CO}$ lie somewhere between $R_\mathrm{sub}$ and $R_\mathrm{trans}$ for all disks.  An important caveat is that, as discussed in Section \ref{sec:modelresults}, $R_\mathrm{CO}$, the upper limit to $R_\mathrm{CO}$ is not well constrained for HD 135344 B,  HD 141569 A and SR 21.  The emission in SR 21 and HD 141569 A could originate at or near $R_\mathrm{trans}$, as is observed for the transitional disk HD 100546 \citep{Brittain09}; however, for HD 135344 B, some of the CO gas must originate from within the inner, optically thin region \citep{Pontoppidan08}.

Using the procedures described in Section \ref{sec:procedure}, we derive significantly larger radii for transitional disks than for classical disks at the same luminosity.  
This difference is typically an order of magnitude --- much larger than our uncertainties --- and holds true for almost three orders of magnitude in luminosity.
Here we discuss some possible explanations for this difference.

One possible explanation is that this is the result of some systematic bias in our analysis.  Two possibilities come to mind. First, since $R_\mathrm{CO}$ depends on the line wings, one might derive smaller inner radii by ``fitting the noise'' at large velocities.  However, since transitional disks tend to have lower continuum S/N levels, and lower line/continuum ratios than classical disks \citep{Salyk11}, this would tend to bias one toward deriving larger characteristic velocities and hence {\it smaller} inner radii for transitional disks.   In addition, we find that the transitional disk radius discrepancy holds even if we examine $R_\mathrm{in}$ derived from the HWHM.   A second possibility is that $R_\mathrm{CO}$ appears to depend somewhat on the choice of lines included in the line composite, with lower-excitation lines yielding larger radii by up to a factor of $\sim2$. Since 6/8 transitional disks in our survey (GM Aur, HD 135344 B, HD 141569 A, SR 21, TW Hya, and UX Tau A) did not have high-excitation emission lines, their low-excitation lines were analyzed instead, and this could result in spuriously large radii for these disks.  However, this bias produces only up to a factor of two difference in $R_\mathrm{CO}$, while the observed difference is an order of magnitude.  Furthermore, the discrepancy is also seen for DoAr 44 and LkH$\alpha$ 330, which do have high-excitation emission lines.  Instead, we suggest that the lack of high-excitation emission is a reflection of the relatively larger emitting radii (and thus lower emitting temperatures) for these disks.  

A physically-motivated explanation for the radius discrepancy is that $R_\mathrm{CO}$ in transitional disks is not set by sublimation, but rather by dynamical truncation by an embedded protoplanet.  Although embedded protoplanets have been posited as a possible explanation for the inner clearings in transitional disks, they are expected to orbit at radii capable of producing the sharp transition between the optically thin inner and optically thick outer disk regions.  Since the $R_\mathrm{CO}$ we find here is usually significantly smaller than $R_\mathrm{trans}$ we would actually need to invoke dynamical truncation by {\it additional} planets, in smaller orbits.  In particular, we would predict planets located at radii near $\sim0.5\times R_\mathrm{CO}$ \citep{Artymowicz94}.  This is in line with recent work suggesting that multiplanet systems may be required to explain transitional disks \citep{Zhu11, Dodson-Robinson11}.  However, this explanation would need to be consistent with the apparent increase in $R_\mathrm{CO}$ with luminosity.   This is not unreasonable, as the mass and location of the protoplanets may depend on the disk and/or stellar mass.  

Another possible explanation for the radius discrepancy, and one qualitatively consistent with the increase of $R_\mathrm{CO}$ with luminosity, could be that
the dust disk is still truncated at $R_\mathrm{sub}$, but that photodissociating UV photons penetrate a finite distance into the tenuous transitional dust disk, making 
$R_\mathrm{CO}>R_\mathrm{sub}$.  A possible quantitative test for this hypothesis would be to calculate a UV penetration depth and compare to the radius discrepancy. However, this is difficult to test in practice, since the local dust density at $R_\mathrm{sub}$ is not known.  (Although SED models can place some constraints on the location and surface density
of the inner disk dust, they do not do so to the precision required here.)  Yet, with sufficiently low dust densities, this explanation is plausible; for example, with $\kappa\sim10^4\ \mathrm{cm}^2\ \mathrm{g}^{-1}$ and a dust density $\rho=10^{-18}\ \mathrm{g\ cm}^{-3}$ (equivalent to a vertical dust surface density of $\sim10^{-6}\ $g cm$^{-2}$ for a scale height of 0.1 AU), the penetration depth (to $\tau=1$) is $\sim7$ AU.  For comparison, \citet{Eisner06} measure a dust surface density of $6.3\times 10^{-7}$ g cm$^{-2}$ at the inner edge of the TW Hya disk.  

A final possibility we consider is that the CO temperature structure in the emitting layer is different for classical and transitional disks.  Since we cannot
distinguish between models with steeply declining line luminosity, $L_\mathrm{CO}(R)$, beginning at $R_\mathrm{CO}$, and models with flat $L_\mathrm{CO}(R)$ at smaller radii 
(see the discussion in Section \ref{sec:modelresults}), it is possible that some aspects of the latter model are more appropriate for transitional disks.  This possibility could be tested
with radiative transfer code that properly treats the gas heating and cooling as well as the line radiative transfer.  

It is interesting to note that \citet{Pontoppidan11}, in contrast, find no significant difference between SA radii for transitional and classical disks.  As discussed in Section \ref{sec:sa}, the SA radii are complementary to $R_\mathrm{in}$, as they represent the flux-weighted mean radius.  Therefore, the results of \citet{Pontoppidan11} suggest that differences in CO emission between transitional and classical disks are erased at large radii, and the outer extent of the emission is similar for the two classes of disks.  If the outer extent is the same, but the inner radius is larger for transitional disks, the emission from these disks should come from a smaller range of radii.

\subsection{Inner Radii and Results from Line Flux Models}
\begin{figure}
\plotone{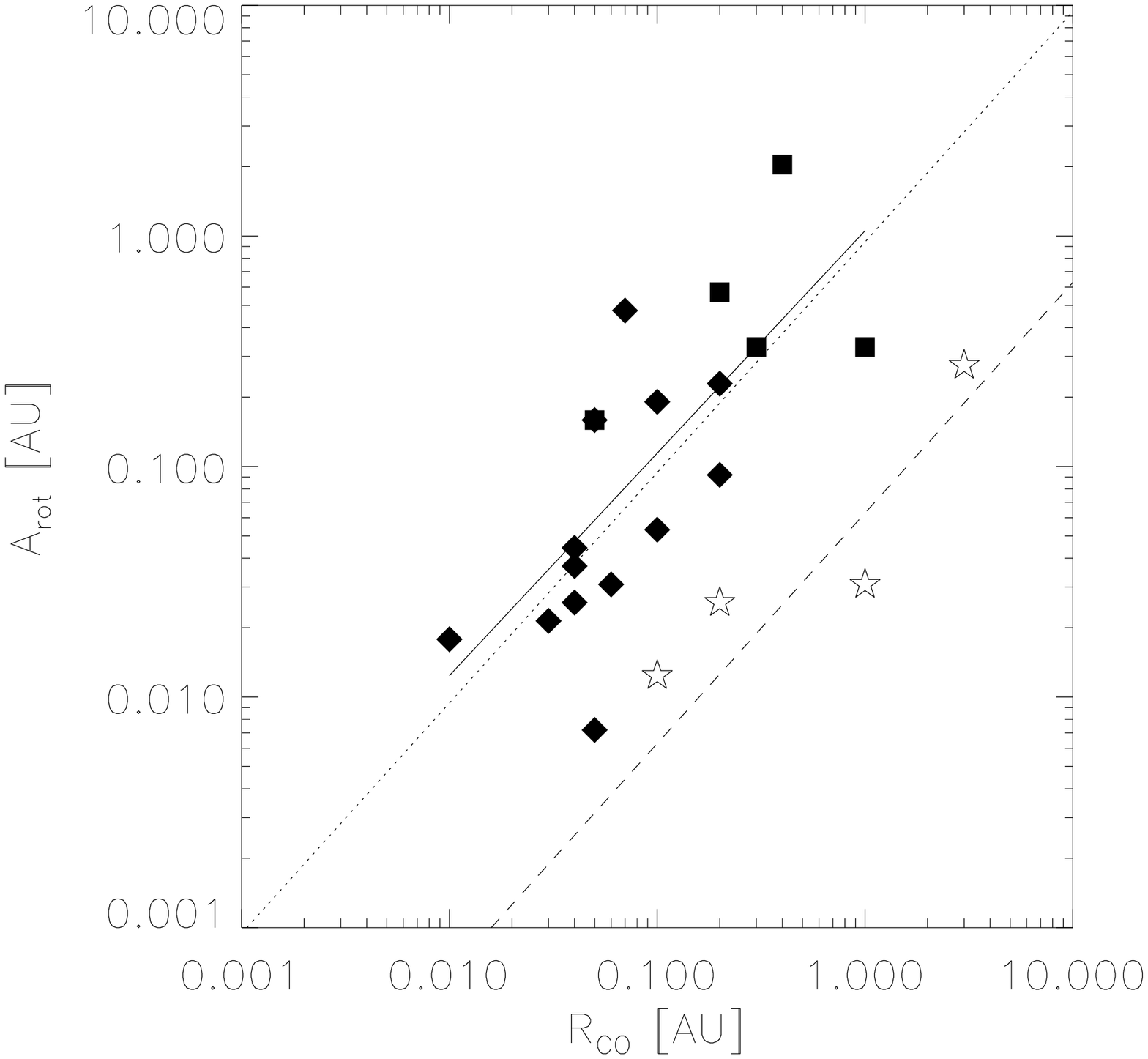}
\caption{Plot of best-fit model area against CO inner radius.  Symbols are the same as in Figure \ref{fig:lrin_acc}.  The solid line shows a linear fit to non-transitional sources; dashed lines correspond to $A=2\pi R\Delta R$ with $\Delta R=0.15$ (top) and $\Delta R=0.01$ (bottom).   (HD 141569 A, not shown, has an anomalously high, and probably incorrect, emitting area.)
 \label{fig:larea_plot}}
\end{figure}

If the CO line flux is dominated by emission from near $R_\mathrm{CO}$, we should expect a positive correlation between $R_\mathrm{CO}$ and the characteristic
emitting area, $A$, derived from the rotation diagram fits.  In fact, as shown in Figure \ref{fig:larea_plot}, we find that for classical disks, $A$ is approximately $\propto R_\mathrm{CO}$.  If the emission comes from a ring located at $R_\mathrm{CO}$ and of thickness $\Delta R$, then the rotation diagram results are consistent with $\Delta R=0.15$ AU.  Interestingly, the relationship is quite different for transitional disks.  Excluding HD 141569 A, which has an anomalously large area (probably due to poorly constrained high-$J$ line fluxes) a similar calculation yields $\Delta R=0.01$ AU.  This is at least qualitatively consistent with the fact that the emission from transitional disks may come from a smaller range of radii, as suggested by $R_\mathrm{in}$ and $R_\mathrm{SA}$.

\begin{figure}
\plotone{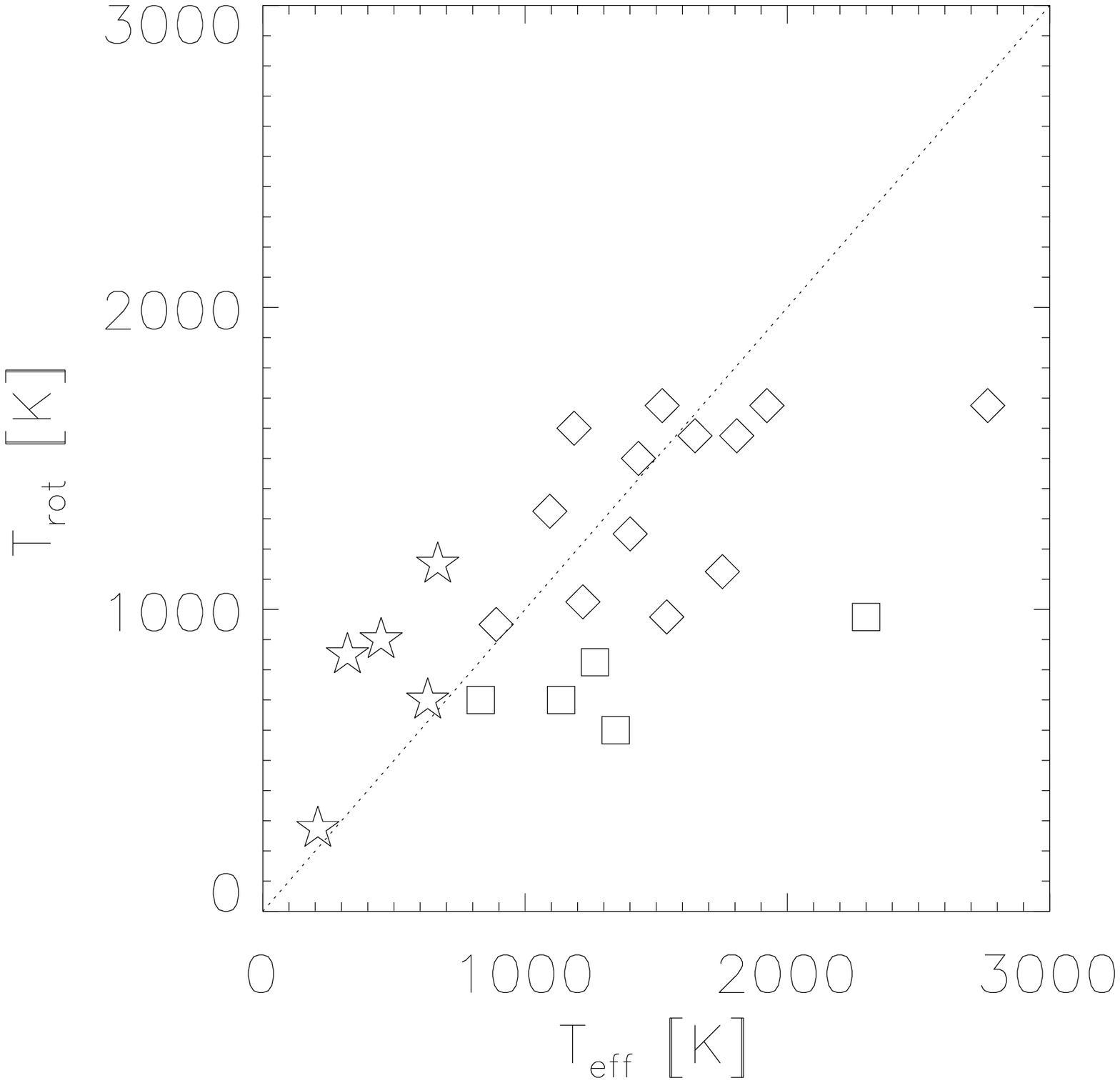}
\caption{Rotation temperature against effective temperature for a blackbody grain fully exposed to stellar and accretion radiation.  The dotted line represents
a 1:1 correlation. Symbols are the same as in Figure \ref{fig:lrin_acc}. \label{fig:trot_teff}}
\end{figure}

With a rotational temperature, $T_\mathrm{rot}$, and a location, $R_\mathrm{CO}$, we can explore the relationship between dust and gas temperatures.  In Figure \ref{fig:trot_teff}, we show $T_\mathrm{rot}$ and $T_\mathrm{eff}$, defined as the effective temperature for a blackbody grain fully exposed to stellar plus accretion
radiation.  We find no statistically significant correlation between these two variables.  In addition, while models predict gas temperatures in the upper layers of
disk atmospheres to be in excess of the thermal dust temperatures \citep{Glassgold01, Kamp04}, we find that $T_\mathrm{rot}$ is similar to, or often less than, 
$T_\mathrm{eff}$.  Although we have assumed blackbody grains, a correction for the relative absorption and emission efficiencies for disk grains would make dust temperatures even higher \citep[e.g.,][]{Monnier02}. The similarity between $T_\mathrm{rot}$ and $T_\mathrm{eff}$ is also curious in light of recent results highlighting the need for a high
gas temperature in order to produce observed H$_2$O emission lines, which have critical densities similar to the CO lines observed here \citep{Meijerink09}.  A possible explanation is that $T_\mathrm{rot}$ represents a characteristic temperature for the emitting region, which is lower than the temperature at the inner rim.

We also find lower $T_\mathrm{rot}$ for transitional and HAeBe disks than for cTTs disks, on average.   This result is consistent with the slightly larger values of
$R_\mathrm{CO}/R_\mathrm{sub}$ observed for HAeBe disks, seen in Figure \ref{fig:lrin_acc}.

\section{Conclusions}
In conclusion, we have used observations of CO emission from a large sample of cTTs, HAeBe and transitional disks to probe inner disk structure. 
Using a conceptually simple parameterization of the CO emission as a function of disk radius, we find that CO inner radii can be robustly determined.  
These inner radii are similar to the stellar corotation radii for low-mass stars, but significantly larger
than stellar corotation radii for disks around HAeBe stars.  We find a strong size--luminosity relationship for CO inner radii, and suggest that CO emission
is truncated at the dust sublimation radius.   Transitional disks are obvious outliers, with CO inner radii typically an order of magnitude larger than
classical disks at the same luminosity, even though the emission typically arises from well within the transition radius at which the disk becomes optically thick.

We also compare CO inner radii with gas characteristics derived from rotation diagram fits.  We find that classical and transitional disk line fluxes
are separately consistent with emission from a single temperature ring of width 0.15 and 0.01 AU, respectively, located at the CO inner radius.  We also
find systematically lower rotational temperatures for transitional disks and disks around HAeBe stars, than for those around T Tauri stars, which is
consistent with observed differences in the ratio of CO inner radius to dust sublimation radius.  Finally, we find rotational temperatures similar to, or
slightly lower than, the expected temperature of blackbody grains located at the CO inner radius.

\acknowledgments
The data presented herein were obtained at the W.M. Keck Observatory, which is operated as a scientific partnership among the California Institute of Technology, the University of California and the National Aeronautics and Space Administration. The Observatory was made possible by the generous financial support of the W.M. Keck Foundation.

\end{document}